\documentclass[sigconf]{acmart}
\usepackage{multirow}
\usepackage{subfigure}
\usepackage{balance} 
\AtBeginDocument{%
  \providecommand\BibTeX{{%
    \normalfont B\kern-0.5em{\scshape i\kern-0.25em b}\kern-0.8em\TeX}}}

\copyrightyear{2021}
\acmYear{2021}
\setcopyright{acmcopyright}
\acmConference[MM '21] {Proceedings of the 29th ACM International Conference on Multimedia}{October 20--24, 2021}{Virtual Event, China.}
\acmBooktitle{Proceedings of the 29th ACM Int'l Conference on Multimedia (MM '21), Oct. 20--24, 2021, Virtual Event, China}
\acmPrice{15.00}
\acmISBN{978-1-4503-8651-7/21/10}
\acmDOI{10.1145/3474085.3475529}

\begin{document}
\fancyhead{}




\title{Actions Speak Louder than Listening: Evaluating Music Style Transfer based on Editing Experience}



\author{Wei-Tsung Lu}
\affiliation{%
  \institution{Institute of Information Science, Academia Sinica}
  \city{Taipei}
  \country{Taiwan}
}


\author{Meng-Hsuan Wu}
\affiliation{%
  \institution{Institute of Information Science, Academia Sinica}
  \city{Taipei}
  \country{Taiwan}
}

\author{Yuh-Ming Chiu}
\affiliation{%
  \institution{KKBOX, Inc.}
  \city{Taipei}
  \country{Taiwan}
}

\author{Li Su}
\affiliation{%
  \institution{Institute of Information Science, Academia Sinica}
  \city{Taipei}
  \country{Taiwan}
}

\renewcommand{\shortauthors}{Trovato and Tobin, et al.}

\begin{abstract}
The subjective evaluation of music generation techniques has been mostly done with questionnaire-based listening tests while ignoring the perspectives from music composition, arrangement, and soundtrack editing. In this paper, we propose an editing test to evaluate users' editing experience of music generation models in a systematic way. To do this, we design a new music style transfer model combining the non-chronological inference architecture, autoregressive models and the Transformer, which serves as an improvement from the baseline model on the same style transfer task. Then, we compare the performance of the two models with a conventional listening test and the proposed editing test, in which the quality of generated samples is assessed by the amount of effort (e.g., the number of required keyboard and mouse actions) spent by users to polish a music clip. Results on two target styles indicate that the improvement over the baseline model can be reflected by the editing test quantitatively. Also, the editing test provides profound insights which are not accessible from usual listening tests.  
The major contribution of this paper is the systematic presentation of the editing test and the corresponding insights, while the proposed music style transfer model based on state-of-the-art neural networks represents another contribution. 
\end{abstract}




\maketitle

\section{Introduction}
Automatic music generation is one of the biggest hits in the research of AI and multimedia over the recent years. This wave of technology development has encompassed a wide range of music generation tasks, such as creating original music content (e.g., generating music from random seeds \cite{roberts2018hierarchical}, a few initial notes \cite{huang2018music,sturm2016music} or a genre label \cite{choi2019encoding,dhariwal2020jukebox,huang2020pop}), extending unfinished musical materials (e.g., harmonization \cite{hadjeres2017deepbach, liang2017automatic,yeh2020automatic}, counterpoint \cite{huang2019counterpoint}, drum arrangement \cite{wei2019generating,barnabo2021cycledrums}), and revising known music works (e.g., rearranging accompaniment to imitate another genre \cite{lu2018transferring}). 
Automatic music generation is also a user-centered technology. Each of these techniques corresponds to one specific application scenario for a specific type of target users, who can be music service providers, listeners, producers, or others. User-centered evaluation or, subjective evaluation of music generation models have therefore become an important yet less mentioned topic in the research of music generation.

The subjective evaluation of music generation techniques has been mostly done with questionnaire-based listening tests alone, with typical questions such as `does the music sound good to you?' or other task-related questions about users' listening experience. Other perspectives of user experience, such as musicians' experience of revising computer-generated music samples according to the functional purpose or their aesthetic taste of music, are often ignored.  
Although many studies on music generation set the application scenario as assisting professional musicians in
create music \cite{huang2018music}, very few of them evaluate their studies quentitatively from the perspective of composers, arrangers and music producers. 

In this paper, we propose a systematic evaluation method, which is called an \emph{editing test}, from the perspectives of general music \emph{editors} who use music generation results as a draft or intermediate product for further revision in their workflow of composition or arrangement. In our proposed testing scenario, users revise the generated results, and the loading users need to pay to polish the generated music can be a measure of the generated music sample. More specifically, we assume that the quality of the generated music sample is measured by the number of keyboard/ mouse actions and editing time (these measures will be called \emph{loading metric} in this paper) during the editing process. The quality of a music sample is considered as better if an editor can make the sample satisfactory using less keyboard/ mouse actions or time on a commercial music editing software. Such an editing test is more quantitative in comparison to the questionnaire-based testing process and can also raise new research questions on the user study of music generation. 
To the best of our knowledge, this paper represents the first systematic investigation on evaluating the quality of music generation based on music editing footprints.

To facilitate the editing test, first, we need to specify that the actions made in the editing process should be restricted to technical ones such as removing non-chord notes or adjusting the bass line, rather than adding new and creative ideas. In this case, the tasks of creating original music contents and extending unfinished musical materials might not be suitable for the editing test since it is unreasonable to set a standard goal or a hard restriction of editing actions. On the other hand, the tasks of revising known music works, such as transferring the style of the accompaniment of homophonic music \cite{lu2018transferring} is more suitable since the melody of the music and the target genre are fixed, and the editor's job is to revise the `inappropriate' contents under such conditions. Therefore, we will focus on the \emph{style transfer} task in this paper. However, an additional issue is that having a comparative study of style transfer is hard; style transfer is a broad term and contains a number of subtasks with different conditions such as texture and the number of polyphony. This issue can be solved by designing a new model which performs the same purpose as the baseline model does. 

Therefore, we propose the Bidirectional Music Style Transformer (BMST), a new style transfer model directly modified from \cite{lu2018transferring} and is dedicated to the same task: re-arranging the accompaniment part of a homophonic music piece to fit a target style while fixing the melody content. The systematic study on editing test can therefore be performed on the two models. By answering the research questions raised in Section \ref{subsec:editing_test}, we demonstrate the feasibility of adopting the proposed quantitative approach to evaluate users' editing experience, and also investigate the multi-facet characteristics of editing tests which cannot be observed with listening tests alone. 

\section{Related work}
\label{sec:related}
Music style transfer \cite{dai2018music} has been referred to as the style transfer between various semantic domains, 
such as timbre and instrument transfer in audio \cite{engel2017neural,verma2018neural,lu2018play}, 
performance rendering from symbolic to audio \cite{hawthorne2018transformer,maezawa2018deep}, 
and composition style transfer 
to modify the harmonic, rhythmic or structural attributes of music at the score level  \cite{malik2017neural,brunner2018symbolic,lu2018transferring}, the latest one will be the main focus of this paper. 
Recently, deep learning models have been used in fitting the style of a specific music corpus. Neural networks have been widely investigated in symbolic music style imitation, including BachProp \cite{colombo2018learning}, BachBot \cite{liang2017automatic}, COCONet \cite{huang2019counterpoint}, and recent works based on Transformer autoencoder \cite{choi2019encoding} and supervised style transfer with synthetic data \cite{cifka2019supervised}.

Quantitative evaluation for music generation is still an open problem \cite{yang2020evaluation, ens2020quantifying}. Recent studies on music generation typically evaluate their work by combining objective evaluation and user study. Objective approaches mostly take the prediction accuracy or log-likelihood (LL) \cite{boulanger2012modeling}, or the rhythmic, harmonic, structural similarity between the generated results and the training data \cite{dong2017musegan}. Subjective approaches or user studies are mostly questionnaire-based listening tests, which list listening samples together with designed questions for the users to evaluate the quality of the listening examples. In other words, such subjective tests focus on listeners' response regarding how good the generated music \emph{sounds}, in terms of either the scaled rating (e.g., five-point Likert scale) or preference (i.e., AB test) comparing several models, and the quality is usually assessed by the `number of wins' among these models \cite{hadjeres2017deepbach,huang2018music}. 
On the other hand, quantitative user studies from music editors' view are rarely seen. 
Related studies are mostly non-quantitative expert interviews \cite{vogl2016intelligent, huang2020ai}. In a recent study which evaluates the quality of automatic music transcription method for ethnomusicologists \cite{holzapfel2019automatic}, 
the correlation between the transcription time and user-reported transcription effort is reported, and the results indicate the potential to perform quantitative editing tests for music generation. 
The evaluation of the melody note detection software \texttt{Tony} \cite{mauch2015computer} might be the only one which studied the number of editing actions and editing time as an evaluation criterion. However, its scenario is different from music generation since the music transcription task has a unique ground truth.


\section{Style transfer methods}\label{sec:method}


A comparative study is performed on two style transfer models which are based on similar architecture. The first model, called the baseline model in this paper, is a re-implementation of \cite{lu2018transferring}. The other model, named as the Bidirectional Music Style Transformer (BMST), is a newly proposed model which is improved from \cite{lu2018transferring}. Both the models are based on the same music language formulation and also keep the same inference mechanism. The two major difference of BMST from the baseline model is that BMST replace the forward and backward components, which are two LSTM networks in the baseline model, by two Transformer blocks, also, besides predicting the pitch activation, BMST further models the onset and offset events which formulate the training to be a multi-task learning problem. In this section, we first focus on introducing the proposed BMST model and then we summarize the improvement we make on the baseline model. 

\begin{figure}
    \centering
    \includegraphics[width=\columnwidth, trim={0cm, 0cm, 0cm, 0cm}, clip]{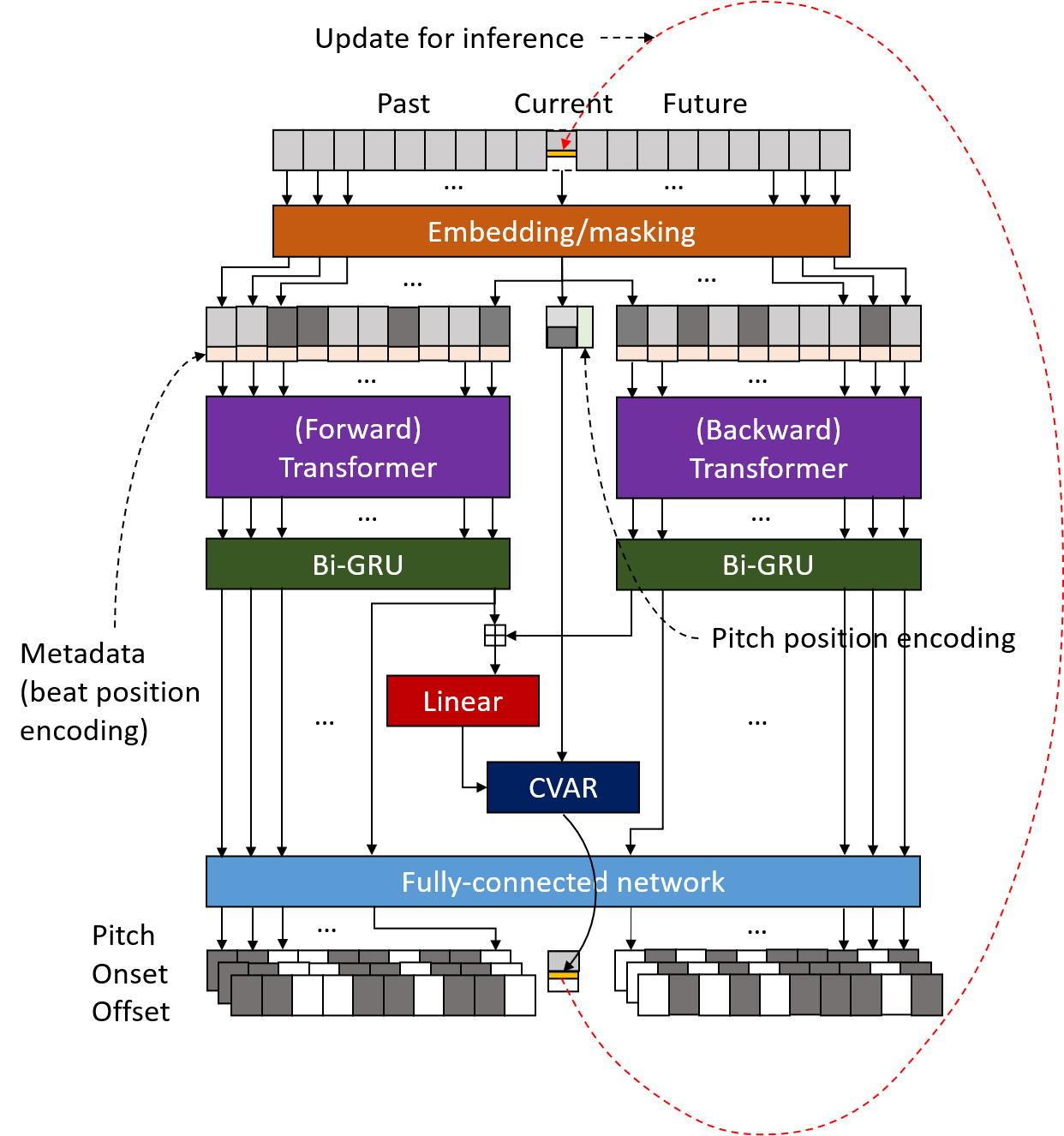}
    \caption{The proposed music style transfer model. Known inputs are in light grey, masks in dark grey, predictions in white, and the updated element is in yellow.}
    \label{fig:system_diagram}
\end{figure}

The architecture of BMST is shown in Figure \ref{fig:system_diagram}. 
In the training stage, a music language model is trained to learn the distribution representing individual music style. The model is composed of two Transformers to model the temporal structure of music, and a convolutional-autoregressive (CVAR) model similar to \cite{lu2018transferring} to capture the pitch-wise conditional probability within defined timestep. In the inference stage, Gibbs sampling is utilized to transfer the style of the input pieces by sampling from the previously learned language model and updating the input piano roll iteratively. 

\subsection{Data representation for polyphonic music}
An input score is represented as a 2-D piano roll $S \in \mathbb{R}^{I \times J}$, where $I=84$ is the number of pitches (from \texttt{A0} to \texttt{G\#7}) and $J$ is the length of the score. 
The minimal timestep is one-eighth of a beat i.e. a 32nd note. In the embedding process, the input at each timestep on $S$ is first transformed into an $I$-channel encoding $S_t \in \mathbb{R}^{I \times I}$, where each channel represents the one-hot encoding of each pitch. All channels are fed into a shared linear embedding layer. Then, the outputs are summed up into a single channel, and are concatenated with a 4-D metadata matrix $M \in \mathbb{R}^{K \times J}$. 
including the start symbol, end symbol, and the beat position encoding. The start/end symbols are both binary-valued, where the start/end of the music are value of one and others are zero. The beat position encoding contains two periodic ramp functions (both are in the range of $[0,1]$), 
one is the position within one measure, and the other is the position within one beat interval. Such representations of time grids provide additional rhythmic guidance to the model. 
For simplicity, two notes activated at identical time and pitch are considered as a single note. Also, it should be noticed that the sustained notes will be seen as repeating notes with the same pitch under this representation. 

\subsection{Proposed model}
The proposed model is formulated as below \cite{lu2018transferring}:
\begin{equation}\label{eq:optim}
    \max_{\theta}\sum_{ij}\log p\left(S_{ij}=1|S_{\backslash ij}, M, \theta\right)\,,
\end{equation}
where $S$ represents the input piano roll, $M$ is the corresponding metadata, $\theta$ is the parameters of the model, and $i$ and $j$ indicate the timestep and pitch respectively. This equation indicates a task to predict the likelihood of a pitch $j$ to be activated at time $i$ conditioned on all other note activations and silence. 

The embedding and metadata encoding of the piano roll are fed into the Transformer model \cite{vaswani2017attention, chen2019harmony}. The transformer used in this work is based on the encoder part in \cite{vaswani2017attention}. As we deal with relatively high-resolution data in music scores (i.e., a 32nd note as the time grid),  
long-term dependency is poorly modeled when using a single Transformer: the Transformer tends to predict the same pitches according to its previous and next timesteps, which do contain identical pitch contents in most cases under this time resolution. 
To avoid this issue, first, we divide the input into three parts: past, current, and future, and use two individual Transformers to independently model the information from only the past or the future. This is similar to the architecture in \cite{hadjeres2017deepbach}: the \emph{forward} Transformer takes only the past input and the \emph{backward} Transformer takes only the future input to predict the current embedding. In this case, the model is guided to learn the long-term dependency of music language. Second, to regularize the model to learn from a more general scheme of sequence prediction, we follow the approach in \cite{devlin2018bert} to further mask out 30\% of the timesteps (the dark gray regions in Fig. \ref{fig:system_diagram}) in the input piano roll randomly.  The Transformer is then trained to predict the contents of the current and all the masked timesteps. To address gradient explosion, which usually happens in the early stage of training, we adopt the pre-layer normalization to stabilize the training of the Transformer \cite{parisotto2019stabilizing}. 


%
Although the attention mechanism in the Transformer has been proven to be powerful at modeling sequential data, one of its weaknesses comparing to the RNN is that the Transformer is unaware of the order of the data, which is however a built-in property for RNN. In the original Transformer model, a sinusoidal-based position encoding is introduced to handle this issue \cite{vaswani2017attention}. By concatenating a unique vector with the input at each timestep, the timesteps with the same input feature are distinguished, and therefore can implicitly provide the information of order. 
Recently, \cite{huang2018music} further proposed to add the \emph{relative position embedding} into the model, and found that the resulting model performs better at producing structural outputs and makes a significant improvement to human perception of the resulting music. In this work, we follow this idea and include the learnable relative position embedding, which corresponds to the relative distances between all the timesteps. The attention unit is formulated as:
\begin{equation}\label{eq:attention}
    \text{Attention}(Q,K,V) := \mathrm{Softmax}\left(\frac{QK^T + S_{r}}{\sqrt{D}}\right)V,
\end{equation}
where $Q$, $K$, $V\in\mathbb{R}^{L\times D}$ are the query, key, and value matrices, which are set the same in the self-attention mechanism \cite{vaswani2017attention}. 
$L$ is the length of the sequence, i.e. the total timesteps of the input given to the transformer, and $D$ is the dimension of the feature spaces. $S_{r}\in \mathbb{R}^{L \times L}$ is the relative position embedding, which is also a trainable matrix \cite{huang2018music}. 

Besides, following the idea in \cite{wang2019language}, we also put a bidirectional Gated Recurrent Unit (bi-GRU) layer after the Transformer modules, which directly model the relation along the time axis. The output of the two bi-GRU networks in BMST are processed in two ways: first, the output positions which related to the current and masked-out timesteps are fed into a shared fully connected network (FCN) to predict the original contents. Furthermore, the model predicts the original contents in three channels: the first is simply the piano roll, the second is note onset, and the last one is note offset. The reason to use additional channels for onset and offset is to encourage the model to learn the concept of sustained notes, since they may be considered the same as repeating notes under the data representation we adopt. Second, the output position related to the current timestep are considered again, concatenated, and then used as the conditional vector for the CVAR model inspired by Wavenet \cite{oord2016wavenet} to predict the harmony structure after a shared linear transform. 


The convolutional-autoregressive model  
describes the harmony structure as the joint distribution among all pitches at a certain timestep. With its autoregressive property, the joint distribution of the piano roll $S_j$ at each timestep $j$ is: 
\begin{equation}\label{eq:autoregressive}
    p(S_{j}) = \prod_{i = 1}^{84}P(S_{ij} | S_{1j}, ...S_{(i - 1)j})\,.
\end{equation}{}

The implementation of Equation (\ref{eq:autoregressive}) using the CVAR model has been seen in various audio and image generation scenarios, such as Wavenet \cite{oord2016wavenet} and PixelCNN \cite{oord2016pixel}. To condition the CVAR model on the previously acquired context information, the outputs from the forward and backward Transformer models are concatenated and then fed into the first layer of the CVAR model at the gated activation unit, which take both the autoregressive and conditioned inputs, as described in \cite{oord2016wavenet}.

Due to the shift-invariance property, using dilated convolution layers in the CVAR model to extract features from piano rolls can be regarded as learning the `relative pitch' rather than the absolute pitch in music. 
To further specify the distribution of the absolute pitch and stabilize the training, we introduce the \emph{pitch position encoding}, where we concatenate the input embedding at the current timestep with a seven-bit binary vector which represents the order of pitch: \texttt{0000000} represents \texttt{A0}, \texttt{0000001} represents \texttt{A\#0}, and so on. Seven bits suffice to represent all pitches in the piano roll.
The pitch position encoding is concatenated to every corresponding pitch and is then fed into the CVAR model as extra channels. 

The proposed model is trained as a multi-task learning problem:
\begin{equation}\label{eq:loss}
    \mathcal{L} = \mathcal{L}_{\text{F-TR}} + \mathcal{L}_{\text{B-TR}} + \lambda \cdot \mathcal{L}_{\text{CVAR}}\,,
\end{equation}
where $\mathcal{L}_{\text{F-TR}}$ and $\mathcal{L}_{\text{B-TR}}$ represent the loss computed from the forward and backward Transformers shown in Fig. \ref{fig:system_diagram}, respectively. Each of the two loss terms 
consists of the prediction error of the pitch/non-pitch events, onset/non-onset, and offset/non-offset events for all the pitch positions within all the masked timesteps. $\mathcal{L}_{\text{CVAR}}$ is the prediction loss of the target pitch at the current timestep. We set $\lambda$=0.5 in this paper.
Since the note onset and offset events are sparse in 
a piano roll representation, we employ focal loss ($\alpha_t = 0.25, \gamma = 2$) \cite{lin2017focal} for all the loss term 
to balance the importance of positive and negative samples in the loss function. 

The final output of music style transfer is obtained from the Gibbs sampling process \cite{hadjeres2017deepbach,lu2018transferring,huang2019counterpoint}. As described in the Algorithm 1 of \cite{lu2018transferring}, for every iteration, every element on the 2-D piano roll matrix $S$ is visited, predicted by the BMST model according to its context, and updated. To keep the original melody content in the style transfer result, we impose the melody score on the piano roll at the beginning of each iteration.  Independent Gibbs sampling is used to further enhance the performance, by using an annealed masking probability $\alpha$ to control the portion of elements that are to be updated independently in the piano roll, in order to reach a stable state in fewer iterations. Given upper and lower bound $\alpha_{\text{max}}$ and $\alpha_{\text{min}}$, the value of $\alpha$ in the iteration $n$ can be computed from the following equation:
\begin{equation}
\alpha_n = \max\left(\alpha_{\min}, \alpha_{\max}-\frac{n(\alpha_{\max}-\alpha_{\min})}{\eta N}\right)\,,
\end{equation}
where $N$ and $\eta$ indicate the total iteration number of Gibbs sampling, and the annealed masking ratio defining the required time for $\alpha$ approaching $\alpha_{\min}$. 
We set $\alpha_{\max}=$ 0.6, $\alpha_{\min}=$ 0 and the total number of iteration $n=15$ in this paper. 
For the generation of onset events, 
we proposed a post-processing procedure by feeding the score after Gibbs sampling to the model again, and select 
the predicted onset channels for use. 
The likelihood of the onset for pitch position $i$ in timestep $j$ is represented in the following equation:
\begin{equation}
    O_{ij} = O_{\text{F-TR},ij} \cdot O_{\text{B-TR},ij}\,,
\end{equation}
where $O_{\text{F-TR},ij}$ and $O_{\text{B-TR},ij}$ are the predicted onsets of forward and backward Transformers at $ij$, respectively. In this work, we consider $O_{ij} > 0.05$ as onset events. 


The BMST model is implemented in Tensorflow. The dimension of the input piano roll is ($J=513$, $I=84$). 
For the Transformer, both the past and future contexts contain $L=257$ timesteps, one current timestep, and 256 contextual timesteps. 
Each of the forward and backward Transformers contains four encoder layers with $D=$ 180 and one bi-GRU layer. 
For the CVAR part, seven dilated convolution layers are stacked in order to cover a pitch range of seven octaves. Each layer contains kernels with the size of two, and the 
dilated rate with 
the power of two depending on the depth. ADAM is used as the optimizer to train the model. The code for BMST and listening samples are available at: https://github.com/s603122001/Bidirectional-Music-Style-Transformer

\subsection{Baseline model}
We re-implement the baseline model following its original design in \cite{lu2018transferring}. Here we list its differences from BMST and also summarize our contributions to the model architecture:

\begin{itemize}
    \item The baseline model uses Long-Short-Term-Memory (LSTM) network to model the temporal structures. We replace it with Transformers in BMST. Transformers has shown state-of-the-art performance in the task of music generation \cite{huang2018music} which make it a more robust choice than the LSTMs.
    
    \item  The baseline model is only trained to predict the defined pitch event in the defined timestep. In BMST, we extend this to be a multi-task learning problem by masking out certain parts of the input piano roll and let the model to predict the missing parts as well. Also, the proposed model is trained to predict the onset and offset events which encourage it to learn better representation for musical structure.
    
    \item The CVAR module in the baseline model only learns the concept of "relative pitch" which shown unstable results in preliminary experiments. To alleviate this issue, additional pitch positional encoding is introduced to BMST which make it learn the ``absolute pitch.''
\end{itemize}

\section{Evaluation methods}\label{sec:exp}

\subsection{Data} \label{subsec:data}
Two datasets are adopted. The first dataset contains 371 Bach’s four-part chorals, which are extracted from the music21 library \cite{cuthbert2010music21}. The second dataset contains 487 Jazz \emph{medium swing} songs collected from iReal Pro.\footnote{https://irealpro.com/} Each Jazz piece has two tracks played in the rhythm of medium swing: piano and bass. During the training process, the pieces are augmented by randomly transposing up or down by at most 12 semitones. Both of the datasets are split into 80 \% training, 10 \% validation and 10\% testing.

We perform style transfer on six music pieces including three kids folk songs (`\emph{London Bridge Is Falling Down},' `\emph{Old MacDonald Had a Farm},' and \emph{Twinkle Twinkle Little Stars}), one classical music pieces (the first 16 measures of Beethoven's \emph{Moonlight Sonata})
, and two rock music pieces (the first 16 measures of Beatles' \emph{Rocky Raccoon} and Radiohead's \emph{Paranoid Android}). The initialized chords of the three kids folk songs are generated by the RNN-based melody harmonization method proposed in \cite{yeh2020automatic}. The generated chords are simply triads in root position and are assigned to the music for each beat. Bach's chorale and Jazz medium swing mentioned previously are taken as the target styles for these music pieces. We, therefore, consider two style transfer tasks: 1) style transfer to Bach's chorale, and 2) style transfer to Jazz. Also, the proposed method and a baseline method are compared. As a result, we have 24 generated samples (6 music pieces $\times$ 2 styles $\times$ 2 models) to be analyzed in the evaluation processes. The 24 samples can be retrieved in the supplementary file.


\subsection{Objective evaluation}
We compare the performance of the baseline and BMST models by computing the element-wise error rate on the 2-D piano roll under a teacher-forcing setting, which means that the predictions are made by conditioning the model on the ground truth data. The prediction error rate for all current timesteps is reported. The music pieces for evaluation are excluded from the training datasets. 
Results in Table \ref{table: objective_acc} indicate that BMST achieves the error rate of 0.1\% for predicting Bach, and the error rate of 4.2\% for predicting Jazz, both of which are much lower than the error rates of the Baseline model. 
We will further discuss that to which extent 
such improvement in error rate is reflected in the subjective evaluation in the subsequent section.

\begin{table}[t]
\centering
\begin{tabular}{|l||c c|} 
 \hline
 Task & Baseline & BMST \\  
 \hline
 To Bach's chorale & 2.5 \% & \textbf{0.1}\% \\ 
 To Jazz medium swing& 16.8\% & \textbf{4.2}\%  \\
 \hline
\end{tabular}
\caption{The element-wise error rate of the music language models used in Baseline and BMST.}
\label{table: objective_acc}
\vspace{-0.5cm}
\end{table}

\begin{figure}[t]
\centering
\subfigure{
\includegraphics[width=.4\linewidth, trim={5cm 0.4cm 4cm 2.3cm}, clip]{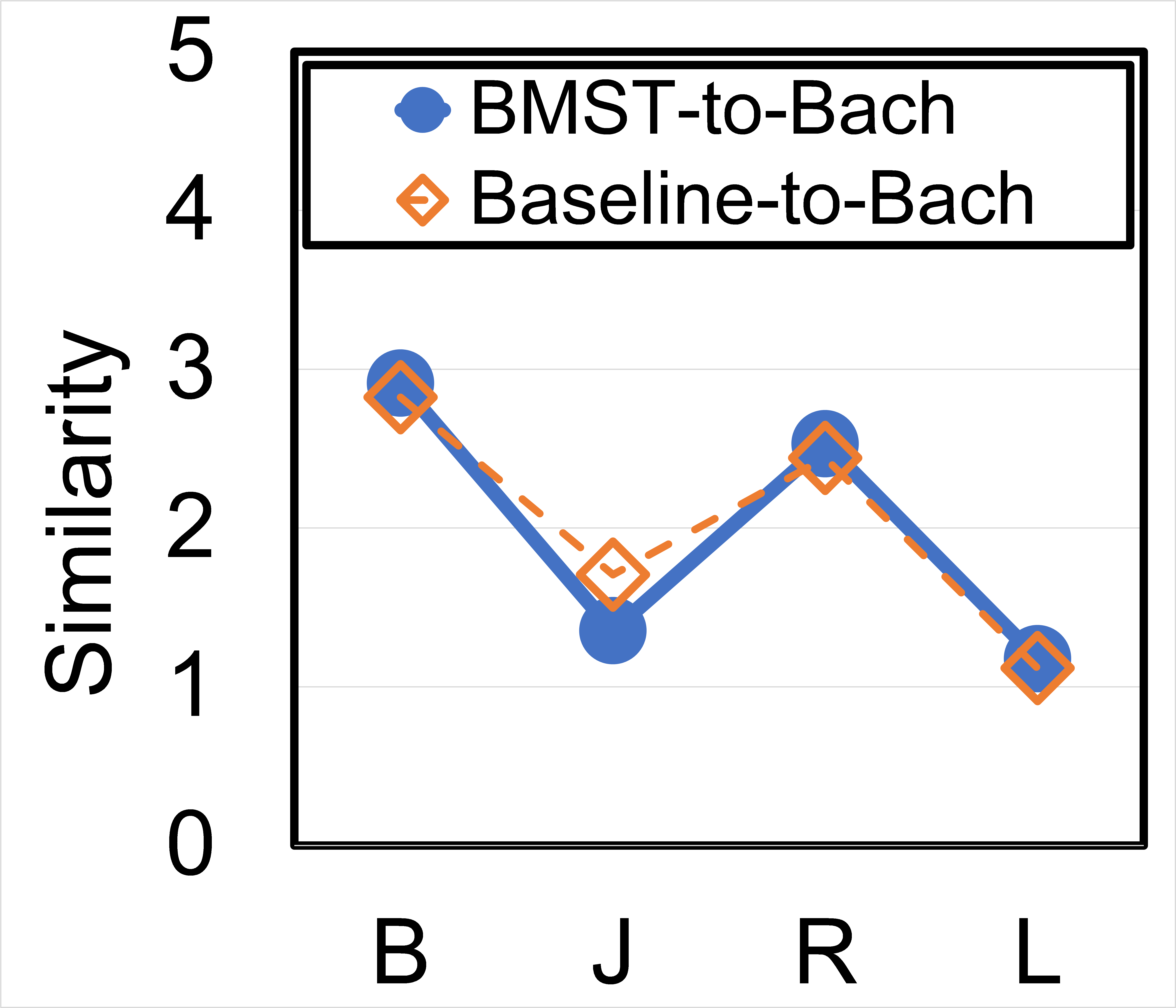}
}
\subfigure{
\includegraphics[width=.4\linewidth, trim={5cm 0.4cm 4cm 2.3cm}, clip]{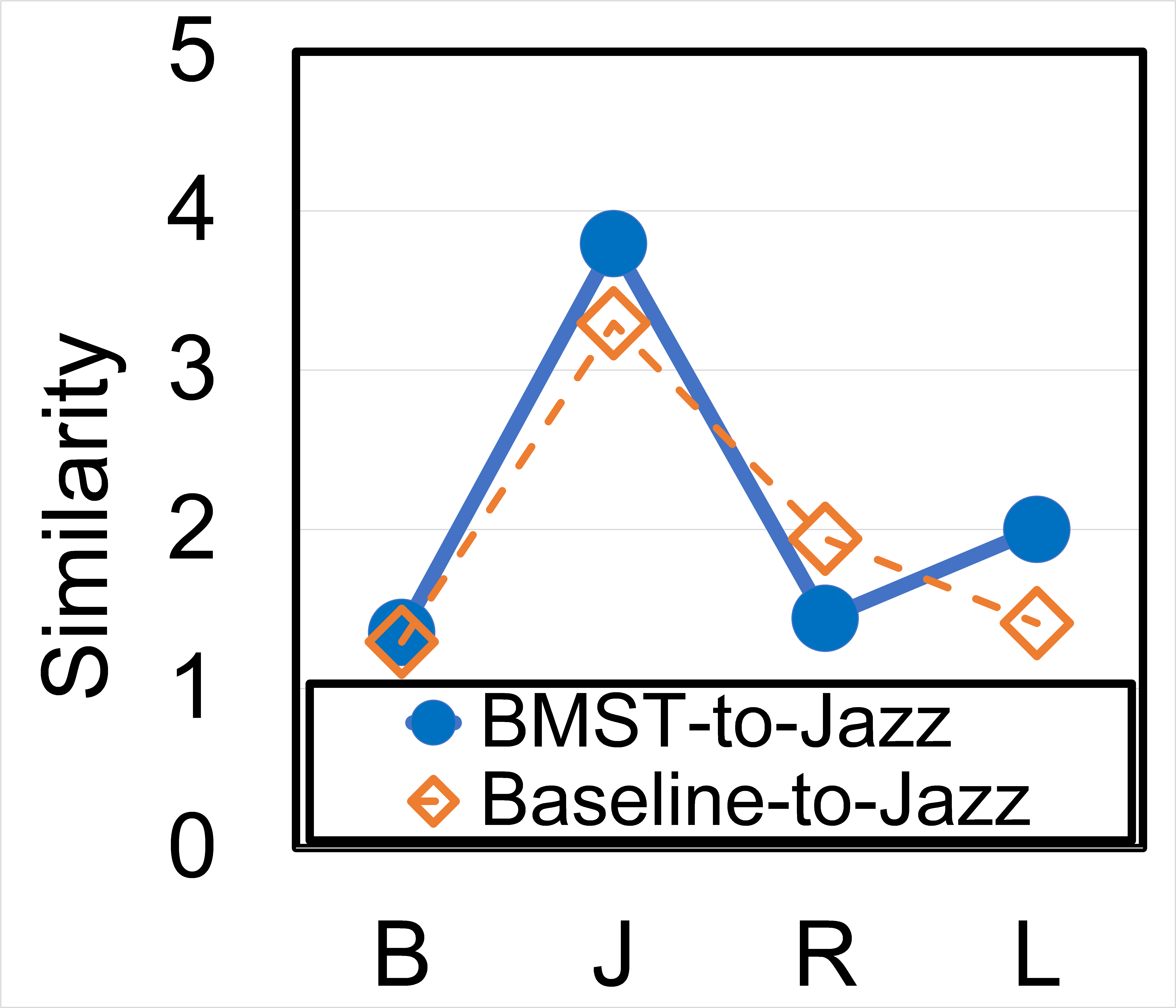}
}
\caption{Result of the listening test. The scores represent the subjects’ evaluation on how similar the style of music to the genre listed in the listening test questionnaire. The four styles in the questions of the listening test are Baroque (B), Jazz (J), Romanticism (R), and Latin (L).}
\label{fig: similarity}
\end{figure}


\subsection{Listening test}
To evaluate the performance of our model from a human perception perspective, a listening test was conducted with 31 participants. 
The 24 generated samples were assigned to three questionnaires. Each participant thus was invited to evaluate eight samples, containing four songs with two style-transferred versions (BMST and baseline). For each transferred version of the music piece, the participants were asked to identify the target style 
from the four styles: Baroque music (i.e., Bach), Jazz, Romanticism, and Latin
, based on their music knowledge and personal perception. The evaluation was on the scale from 1 (low) to 5 (high). Finally, the participants were asked to answer a subjective question: (LQ1) ``Does the generated music sound good?'' The answer was in a five-point Likert scale from 1 to 5 (1 = strongly disagree, 2 = disagree, 3 = neutral, 4 = agree, 5 = strongly agree).

Figure \ref{fig: similarity} shows the average scores on the similarity between the generated results and four target styles rated by the participants. For the cases of style transfer to Bach’s chorale, the rated similarity scores to Baroque and Romanticism are higher than the remaining two styles. 
For the cases of transferring to Jazz, the degree of similarity to Jazz surpasses other types of music. In addition, the overall similarity of BMST is higher than the baseline model in both Bach and Jazz styles. In other words, BMST produces more mimic results for style transfer task, and demonstrates advanced improvement compared to 
the baseline model. 
In addition, the average ratings for LQ1 are listed in Table \ref{table:test_response}, and it is evident that 
BMST outperforms the baseline method in both to-Bach and to-Jazz tasks.


\begin{table}[t]
\centering
\begin{tabular}{|l||c|c|c|c|} 
 \hline
 \multirow{2}{*}{} & \multicolumn{2}{c|}{To-Bach} & \multicolumn{2}{c|}{To-Jazz} \\ 
 \cline{2-5}
  & BMST & Baseline & BMST & Baseline \\  
 \hline
 E-Q1 & \textbf{4.11}\small{$\pm 1.08$} & 3.83\small{$\pm 0.92$} & \textbf{3.33}\small{$\pm 1.08$} & 1.89\small{$\pm 0.83$} \\ 
 E-Q2 & \textbf{3.83}\small{$\pm 0.86$} & 3.61\small{$\pm 0.78$} & \textbf{3.61}\small{$\pm 0.78$} & 2.56\small{$\pm 0.98$} \\
 E-Q3 & \textbf{4.11}\small{$\pm 0.83$} & 4.11\small{$\pm 0.68$} & \textbf{3.61}\small{$\pm 1.04$} & 2.28\small{$\pm 0.96$} \\
 \hline
 L-Q1 & \textbf{3.34}\small{$\pm 0.49$} & 3.18\small{$\pm 0.28$} & \textbf{3.13}\small{$\pm 0.22$} & 2.15\small{$\pm 0.31$} \\
 \hline
\end{tabular}
\caption{The responses on the editing and listening test.}
\label{table:test_response}
\vspace{-0.5cm}
\end{table}

\begin{figure*}

\includegraphics[width=.23\textwidth]{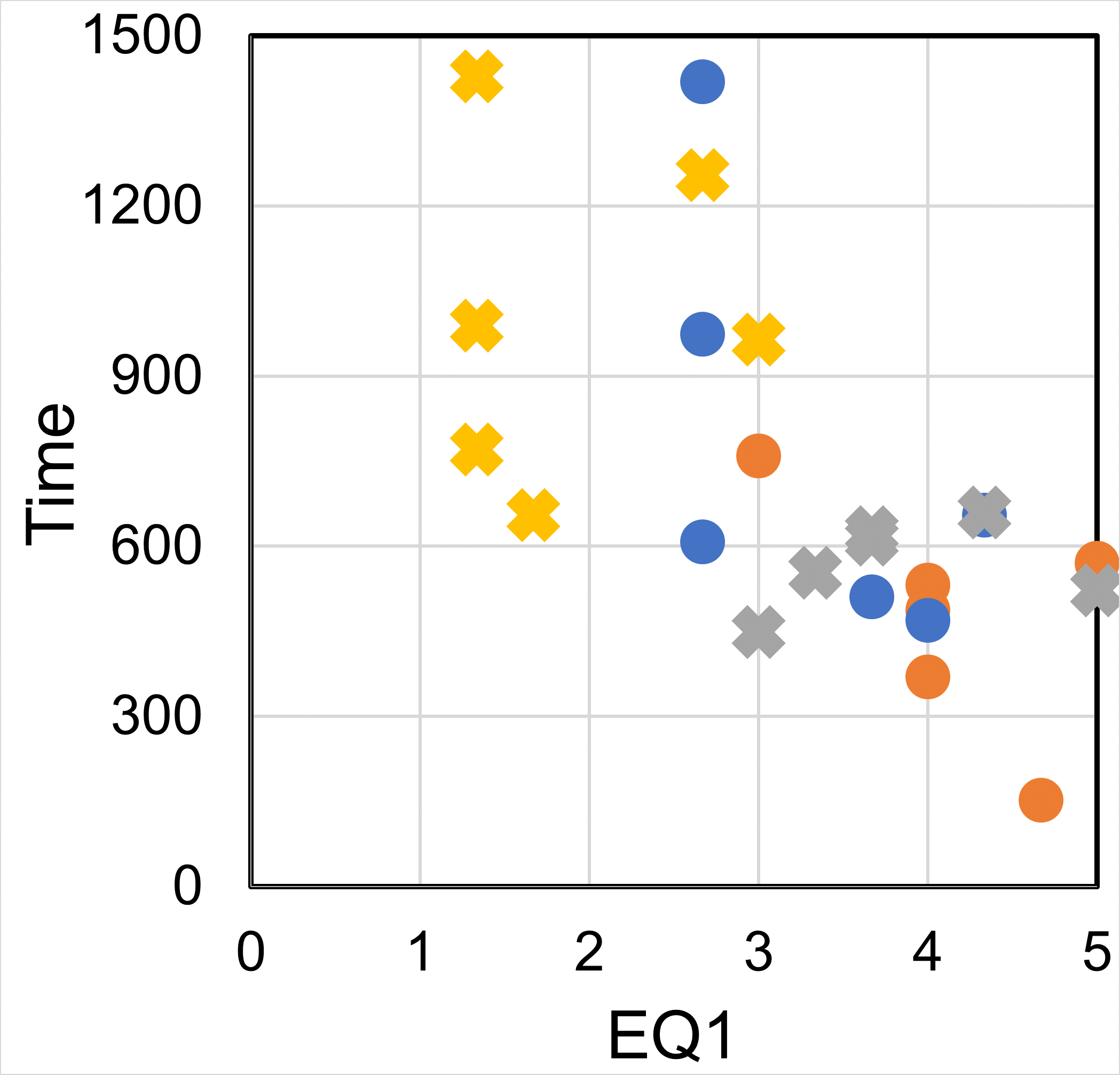}
\hfill
\includegraphics[width=.23\textwidth]{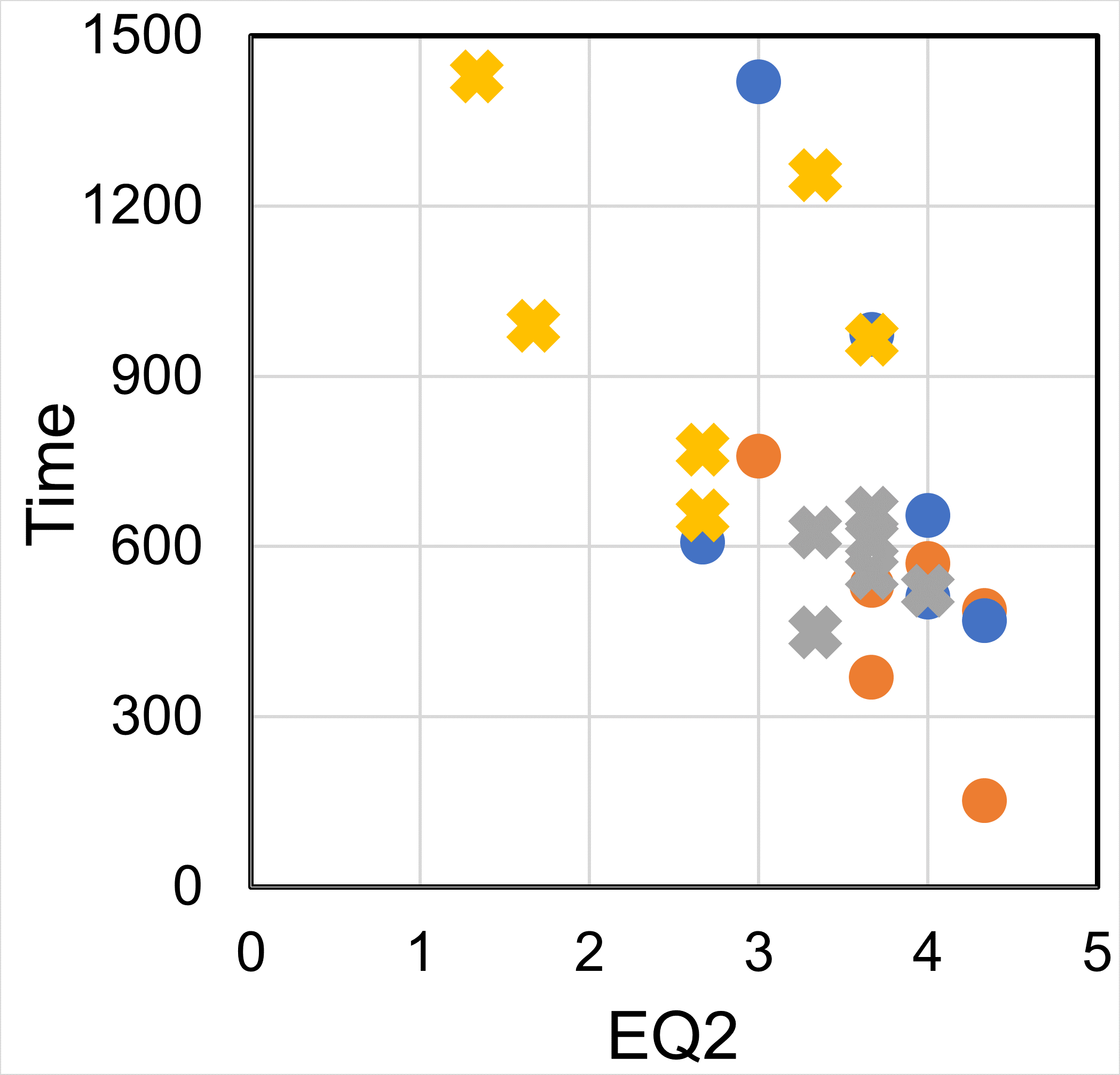}
\hfill
\includegraphics[width=.23\textwidth]{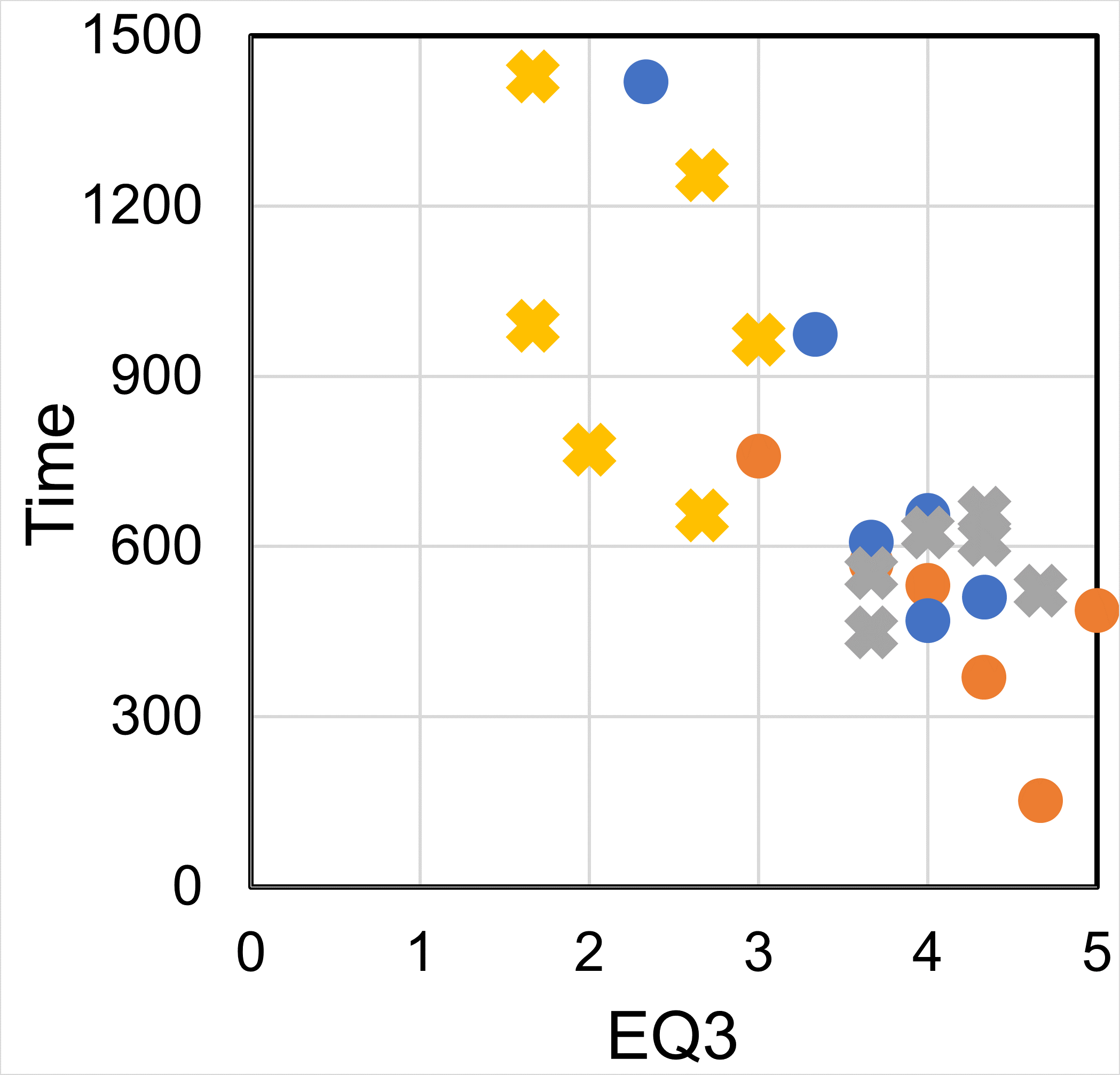}
\hfill
\includegraphics[width=.23\textwidth]{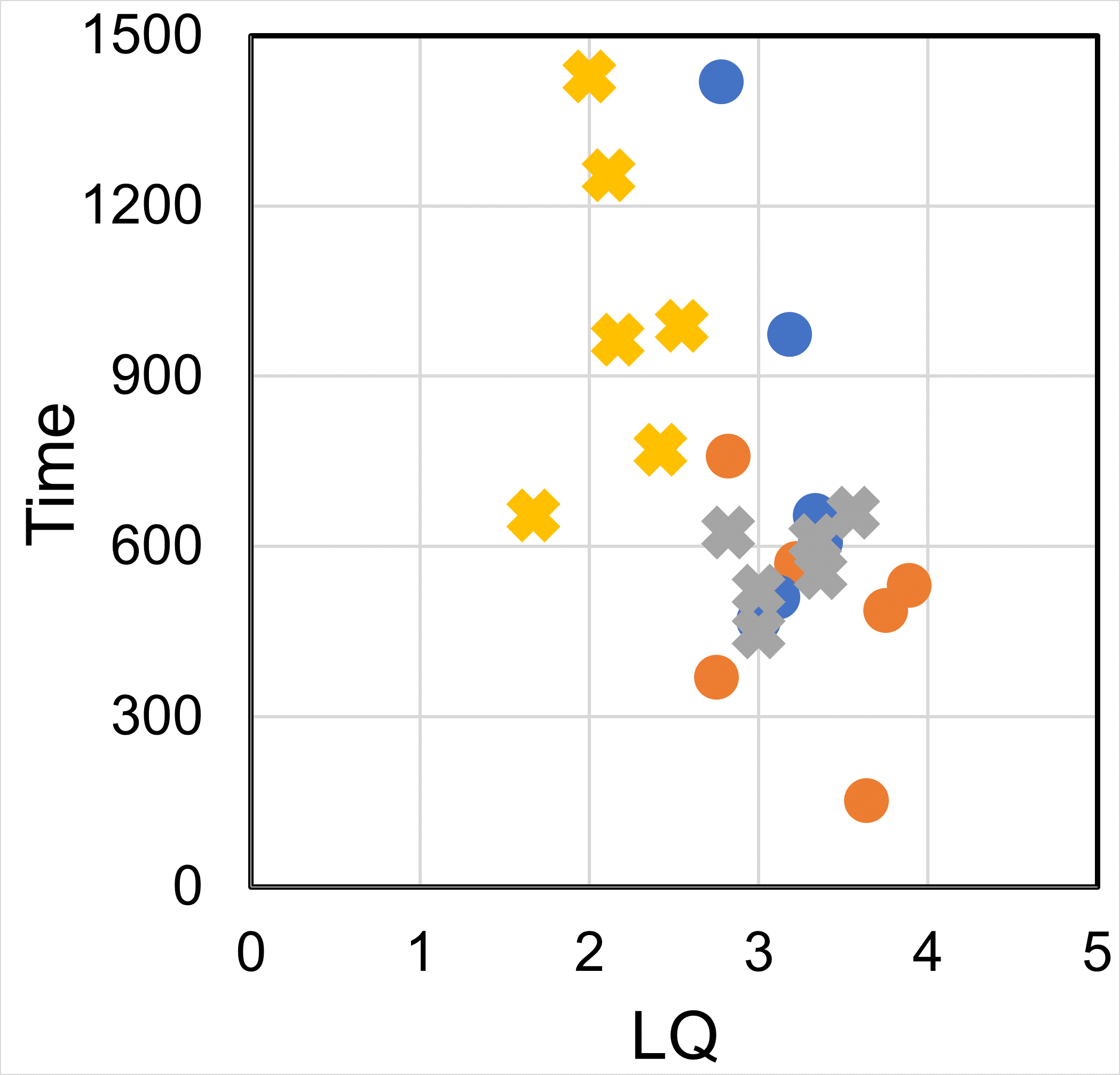}


\includegraphics[width=.23\textwidth]{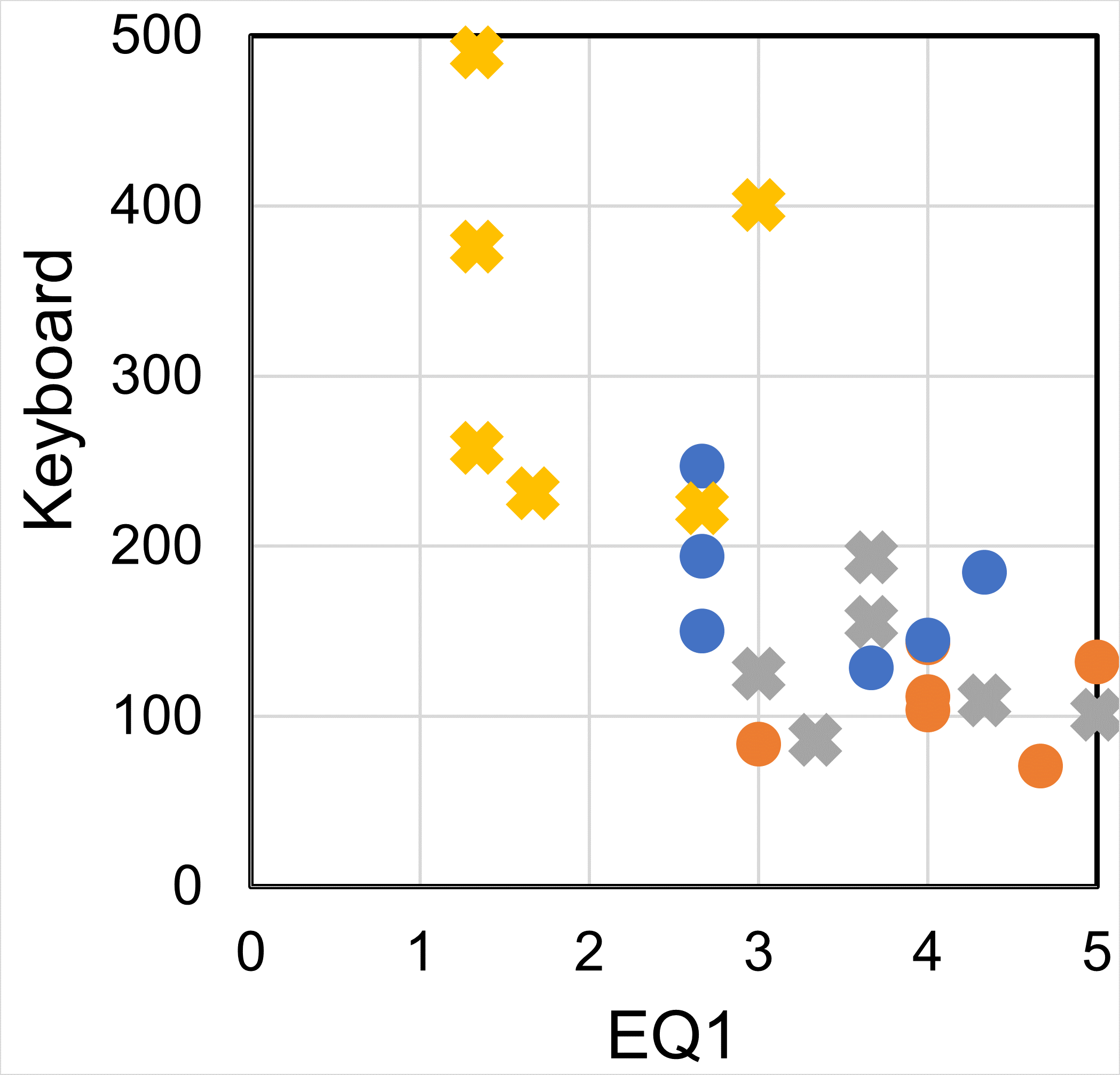}
\hfill
\includegraphics[width=.23\textwidth]{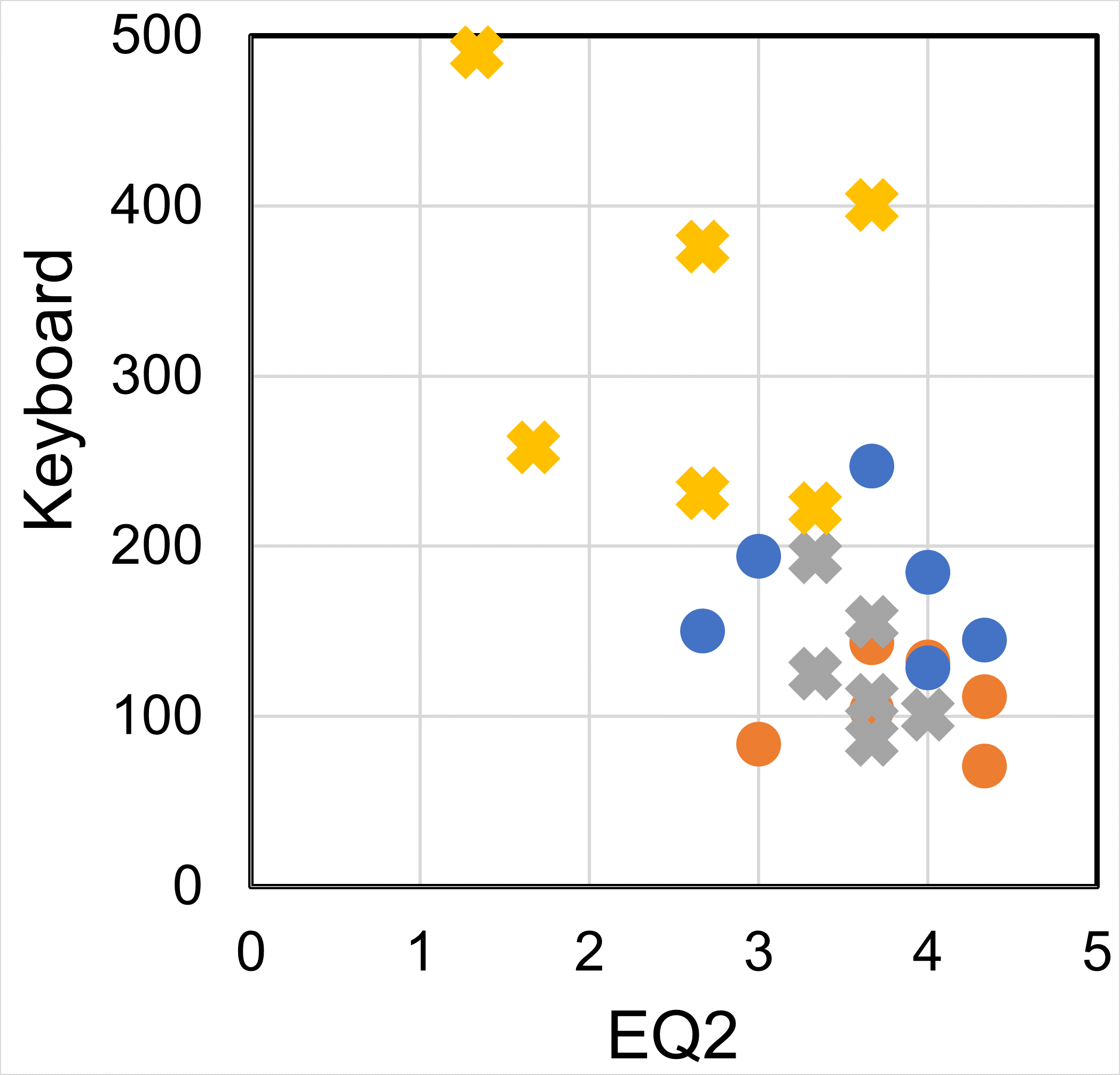}
\hfill
\includegraphics[width=.23\textwidth]{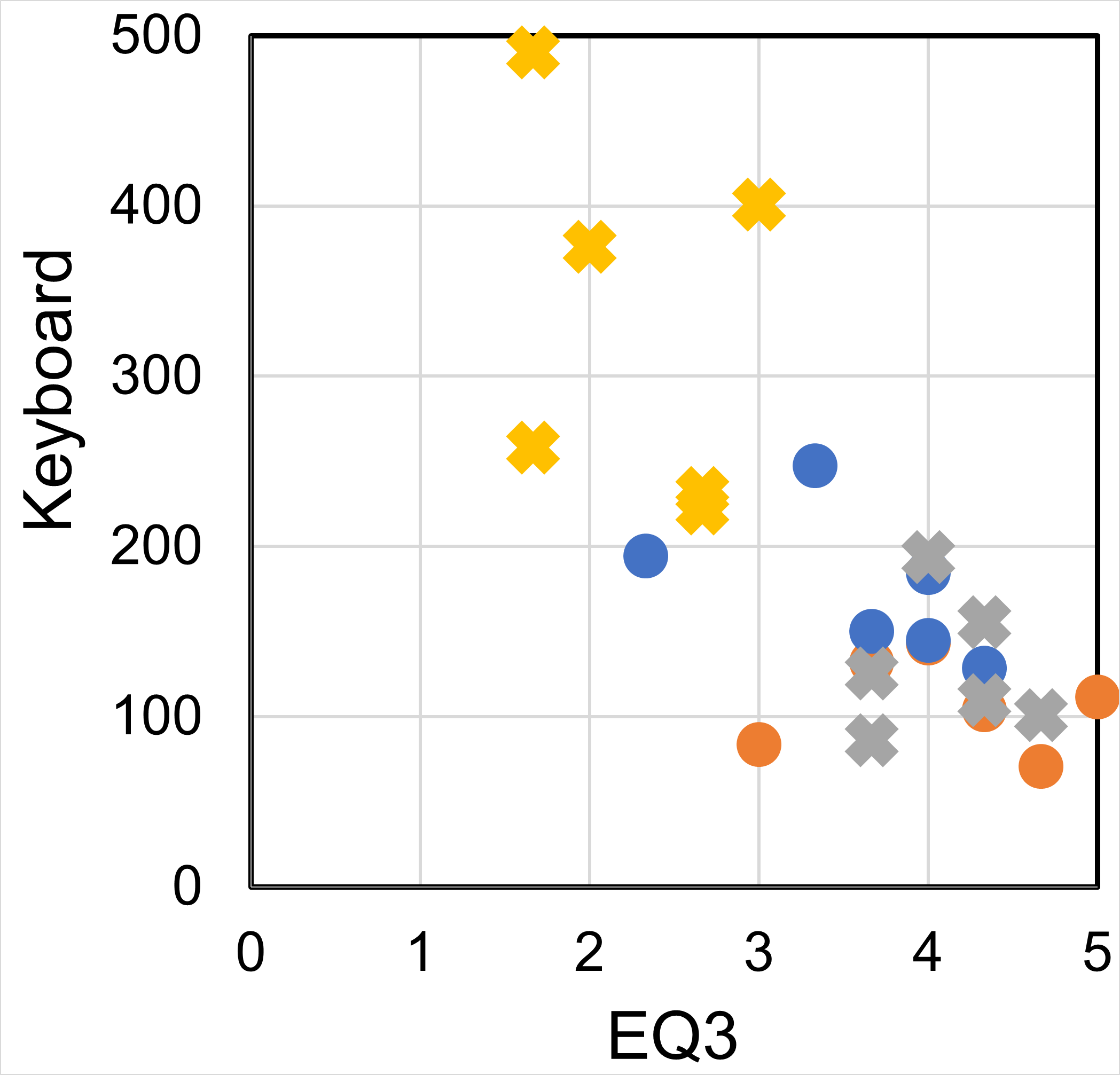}
\hfill
\includegraphics[width=.23\textwidth]{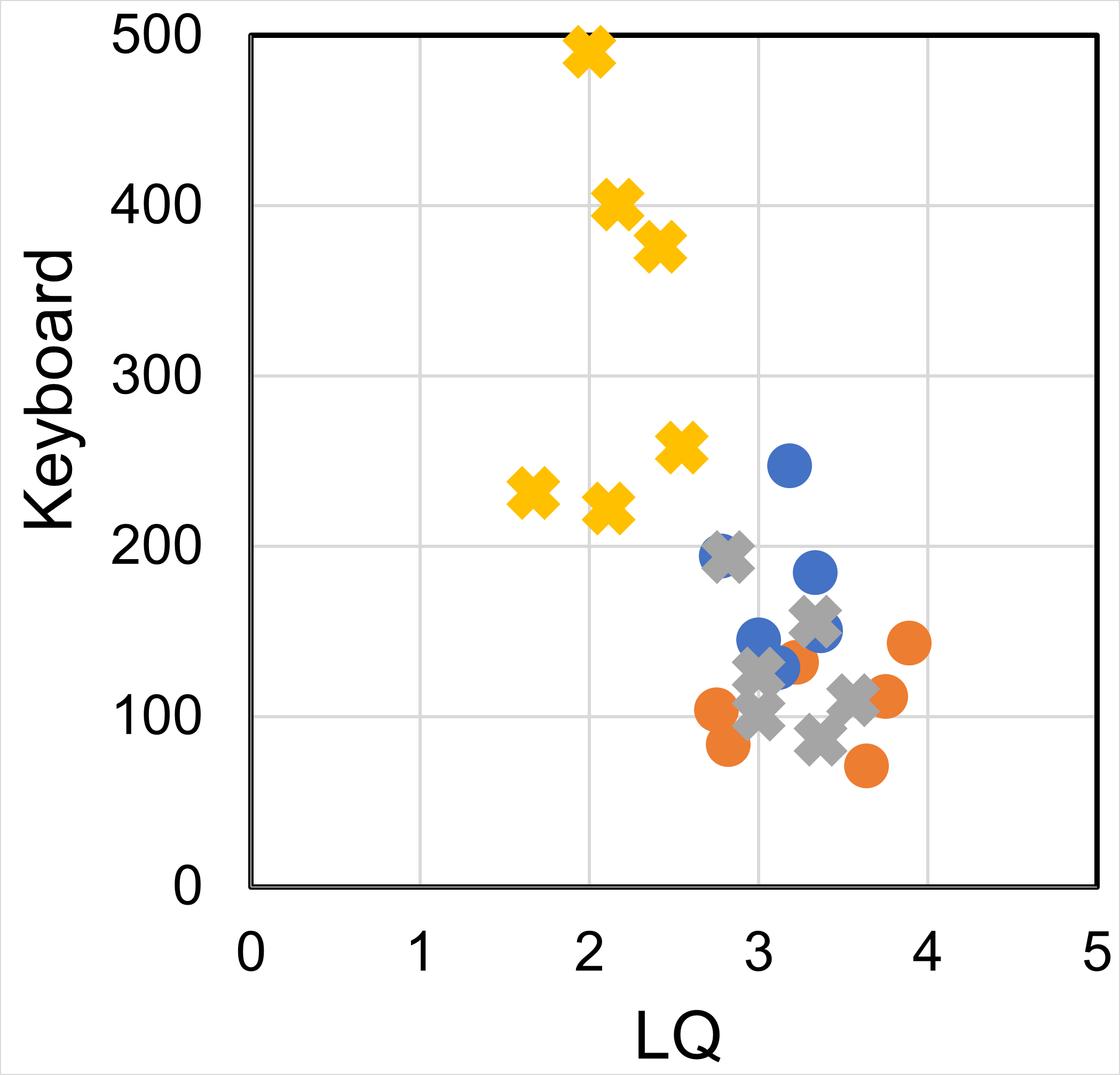}


\includegraphics[width=.23\textwidth]{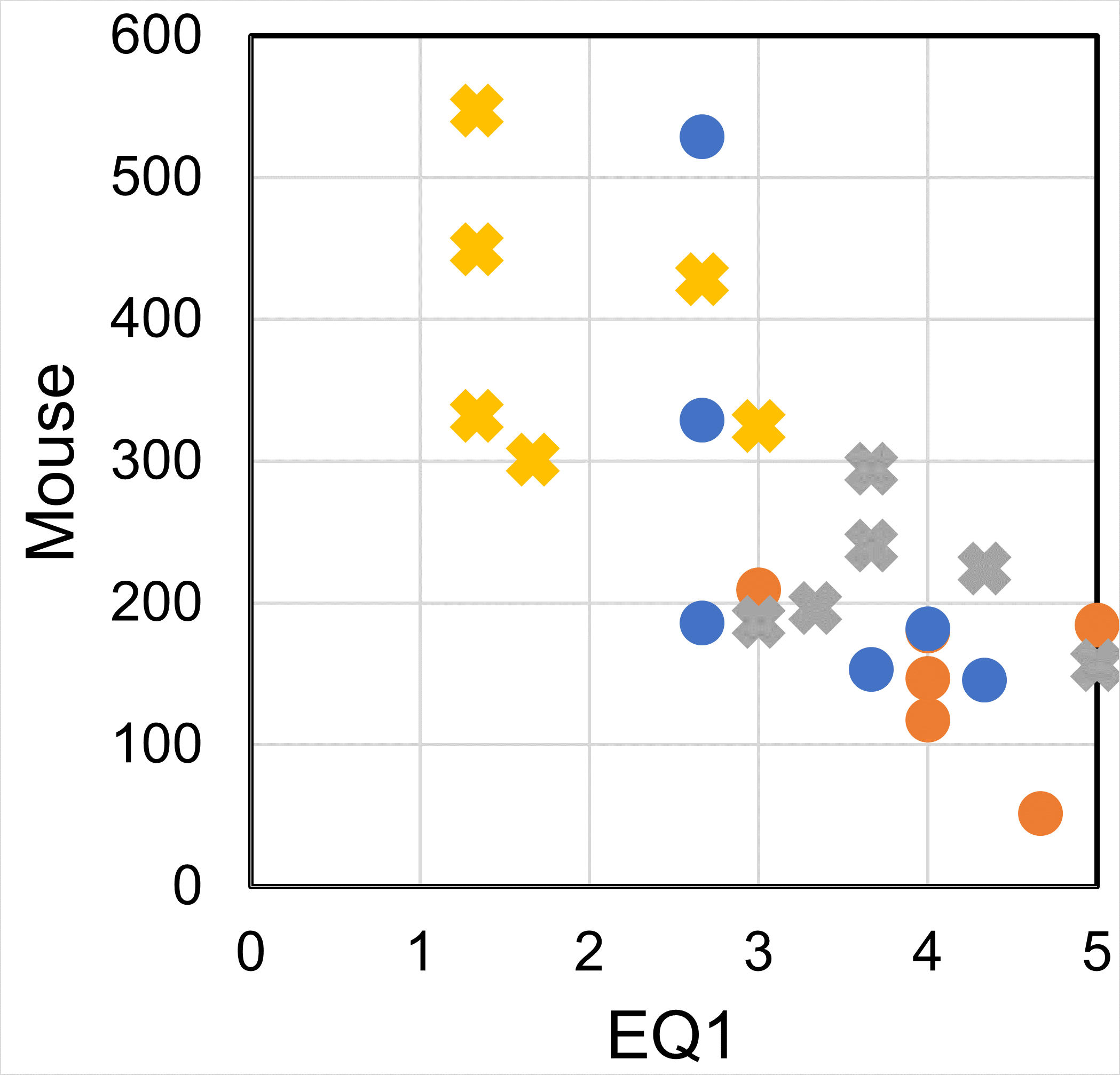}
\hfill
\includegraphics[width=.23\textwidth]{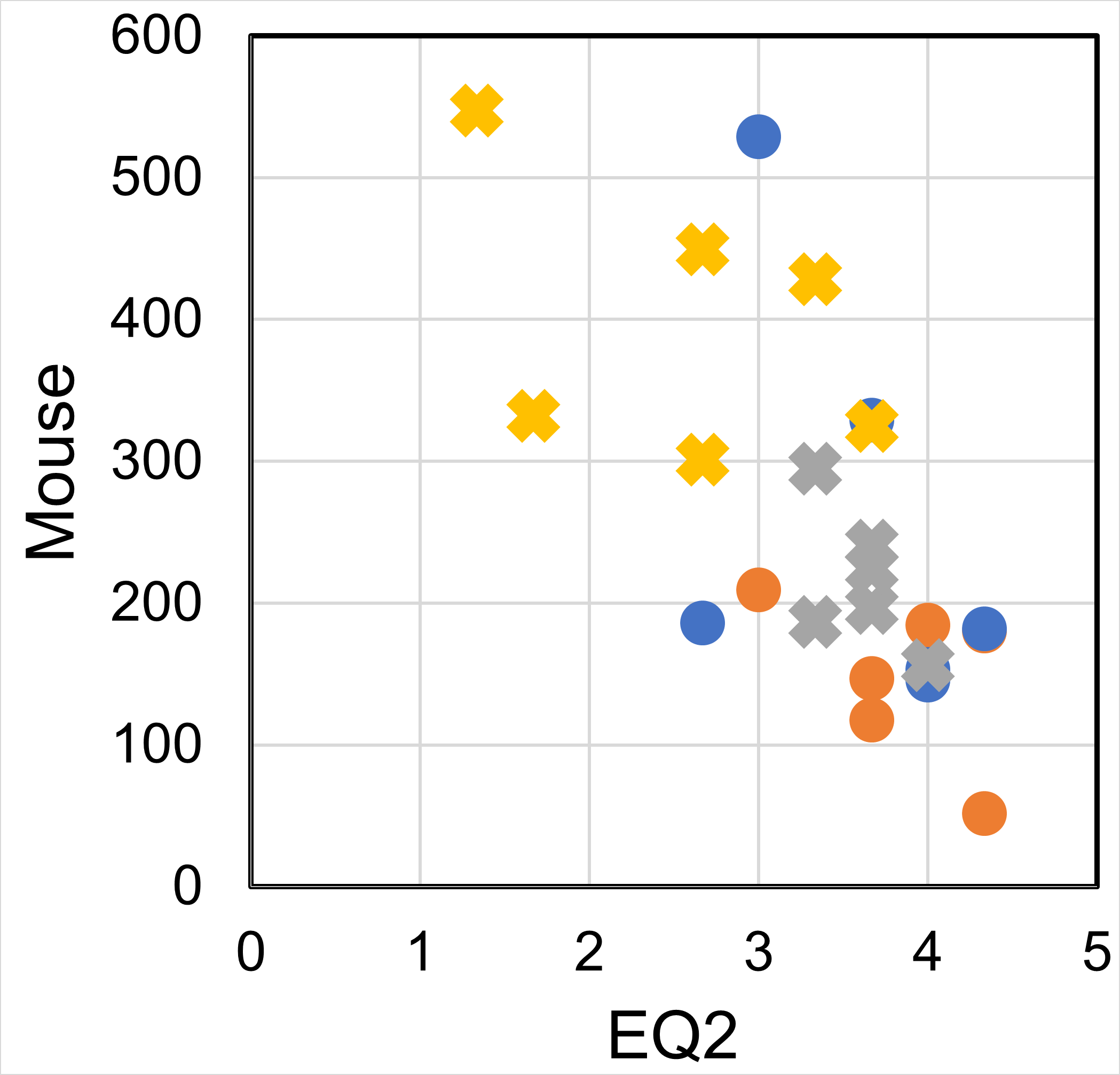}
\hfill
\includegraphics[width=.23\textwidth]{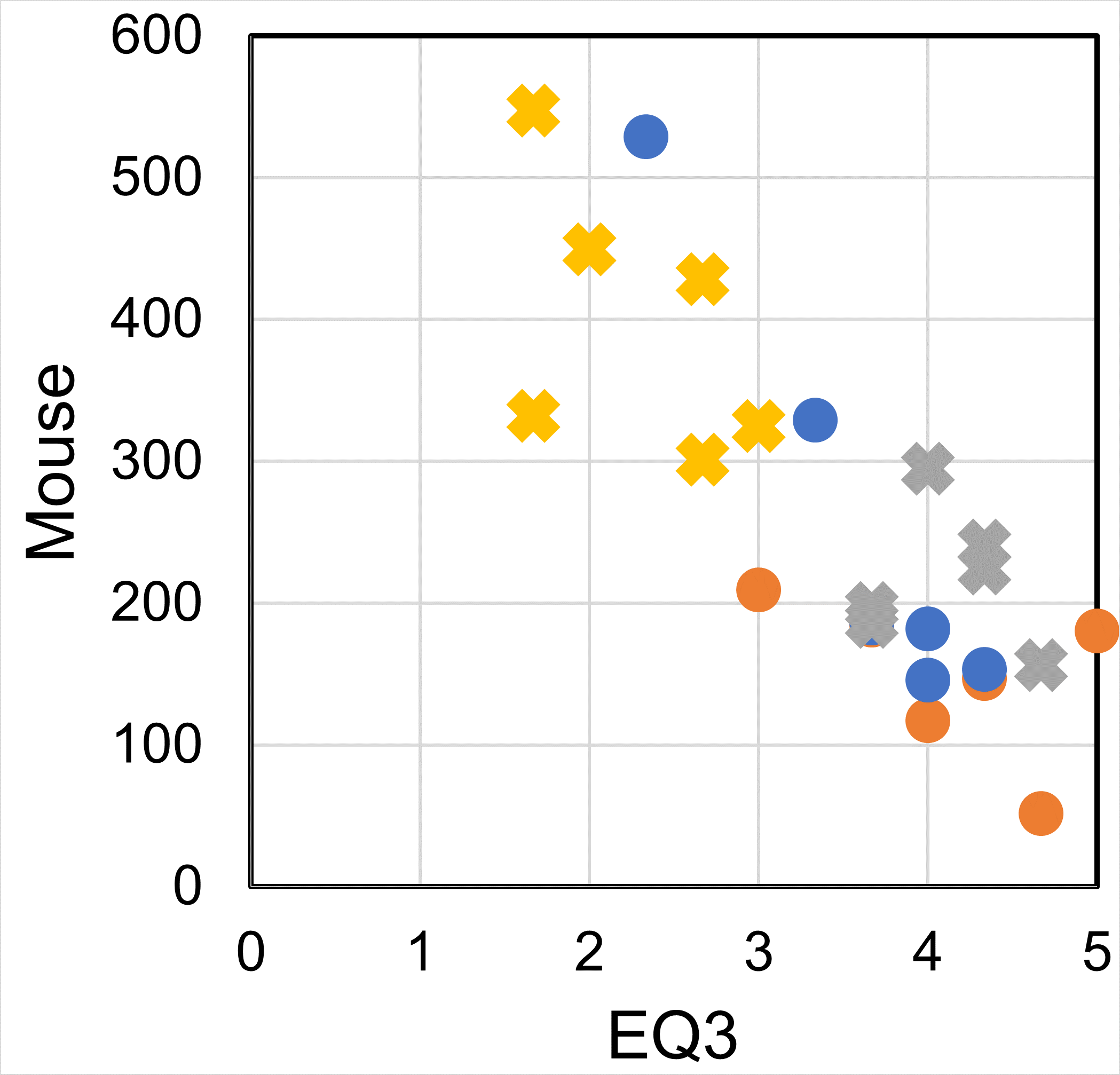}
\hfill
\includegraphics[width=.23\textwidth]{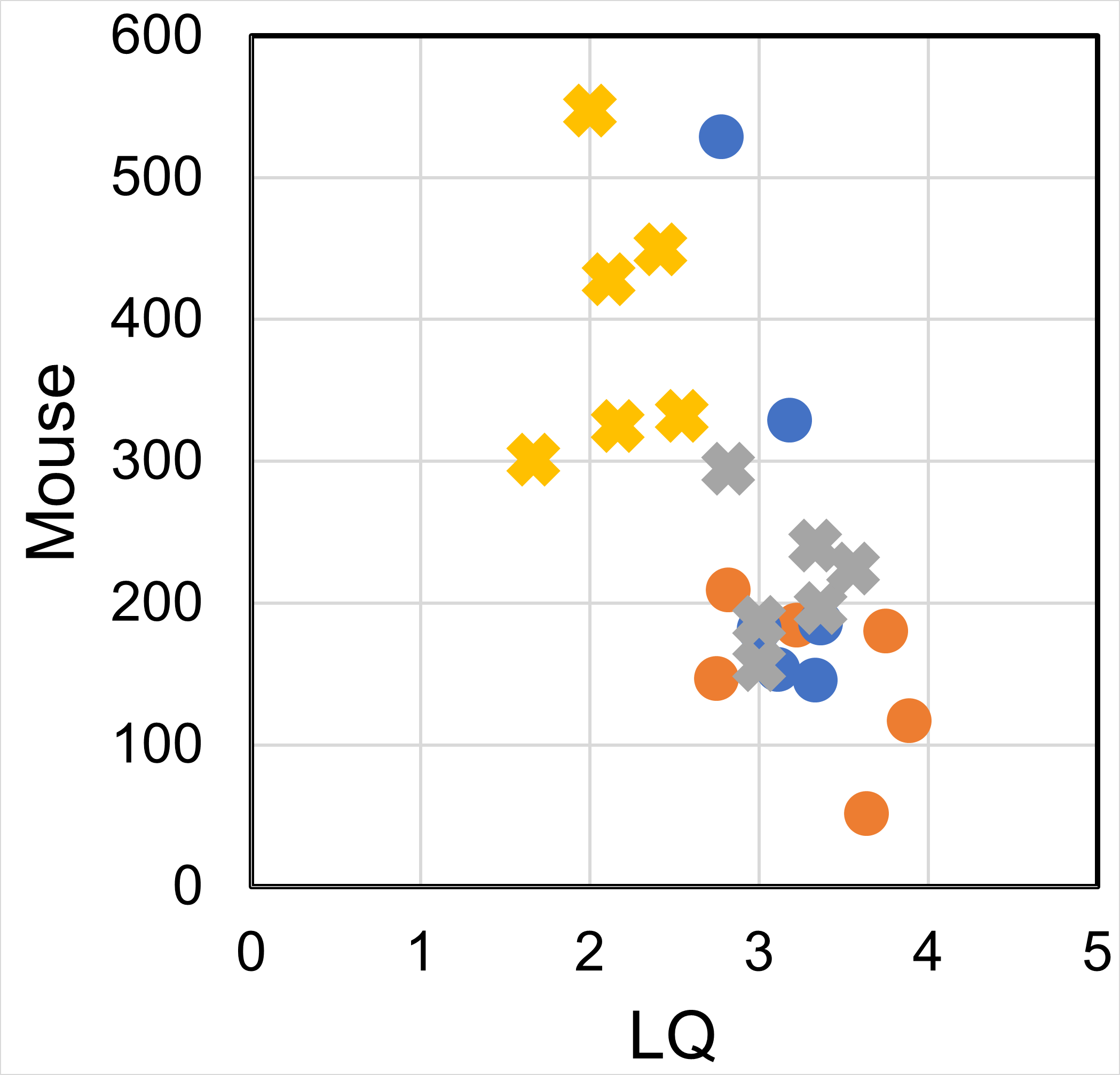}



\caption{The scattering plot of loading metrics and listening/ editing test responses for the 24 samples in the experiments. Orange dot: BMST-to-Bach. Blue dot: BMST-to-Jazz. Gray cross: Baseline-to-Bach. Yellow cross: Baseline-to-Jazz.}\label{fig:editing_test_result}

\end{figure*}

\subsection{Editing test}
\label{subsec:editing_test}
12 professional music editors with substantial musical backgrounds were recruited to edit the pieces generated from both the baseline and BMST models. The participants were asked to `correct' the pieces by moving, splitting, changing the length, or removing the existing notes according to their musical aesthetic perception and their knowledge of musical theory. For instance, they may remove inadequate rhythmic patterns or harmonic progressions. During editing, the participants were instructed to improve the quality of music while avoid adding `new ideas' to the music. That means, we did not encourage the editors to edit the piece according to the target genre; rather, we let the editors decide whether they would like to refer to genre information or not.
We distribute the 24 samples to the editors such that each editor edits six samples and each sample is edited by three times. 

Each editor is allowed to choose the handiest editing software, but is restrained to only use the one they selected throughout the whole process of the experiment. 
The editing for each music piece should be completed within a hard limit of 30 minutes. 
Keyboard and mouse action recording software was installed on the participants' 
computers and was then removed after the experiment was done in order to keep privacy. We consider recording three \emph{loading metrics} during the editing process including: 
1) the editing time, 2) the number of keyboard presses, and 3) the number of mouse clicks. 
For each music piece, the editors also rate their level of agreement on three subjective questions on the feasibility, effectiveness, and usefulness of the style transfer model: (EQ1) I think that this music piece is easy to edit; (EQ2) I think that the music sounds 
good after my editing; and (EQ3) I think that the music clips generated by the model can benefit my process of 
music production. All the questions are in the format of five-point Likert scale. 

\begin{table}[t]
\centering
\begin{tabular}{|l||c c|} 
 \hline
Performance ratio & To Bach & To Jazz \\  
 \hline
 Time & 0.85\small{$\pm 0.37$} & 0.77\small{$\pm 0.26$} \\ 
 Keyboard & 0.92\small{$\pm 0.37$} & 0.59\small{$\pm 0.27$}  \\
 Mouse & 0.72\small{$\pm 0.29$} & 0.65\small{$\pm 0.37$}  \\
 \hline
\end{tabular}
\caption{The performance ratio of BMST over baseline.}
\label{table: performance_ratio}
\vspace{-0.5cm}
\end{table}

Table \ref{table:test_response} summarizes the responses from the participants for questions 
EQ1 to EQ3. For the to-Bach task, both the Baseline and the BMST models receive the average score higher than 3 (neutral) for all questions. For the to-Jazz task, the baseline model was given negative responses (all scores were less than 3), while the BMST model massively improves, which indicates that BMST outperforms the baseline model more on to-Jazz task than to-Bach task. 

We also define the \emph{performance ratio} between the BMST and Baseline methods as the ratio of their recorded loading metrics for the same input music piece. 
A lower performance ratio corresponds to higher quality of the evaluated model since it takes fewer actions. 
Also, the performance ratio is independent from the complexity of the original music content, and therefore serves as a proper metric to 
compare the performance for transferring different music pieces. 
The average performance ratio of all loading metrics for all input music from both BMST and the baseline are shown in Table \ref{table: performance_ratio}. Results of the two target genres (Bach's chorale and Jazz) are both listed. It can be observed from the results that all the performance ratios are smaller than 1, which suggests that editors take less effort to improve the music clips generated from 
BMST, compared to the ones produced by 
the baseline method. 
And also, BMST improves more in the to-Jazz task, since the performance ratios of the to-Jazz task are smaller than the to-Bach task. 

With these positive results in hand, we are further interested in the following research questions (RQ) regarding users' editing experience: (RQ1) Can the loading metrics reflect users' experience of editing? (RQ2) Can the loading metrics effectively measure improvement of users' experience? (RQ3) What is the relationship between different aspects of users' editing experience (e.g., feasibility, effectiveness, and usefulness)? (RQ4) What is the relationship between users' listening and editing experience? These RQs are then discussed one-by-one by a deeper analysis of the results.

\subsubsection{RQ1: Can the loading metrics reflect users' experience of editing?} 
Figure \ref{fig:editing_test_result} shows the editor's responses, listeners' responses, and the loading metrics for the 24 samples. The value of each point (i.e. each sample) is averaged over all the editors'/ listeners' responses. We can observe that first, the samples with low EQ/LQ responses tend to have high loading, which fits our assumption. In addition, we also observe that the samples with low EQ/LQ responses tend to have high \emph{variation} of loading. To discuss the results quantitatively, we compute the Pearson's $r$ coefficients between the editors' rating scores on the rating scores of the three EQs and the values of loading metrics (i.e., time and keyboard/ mouse actions). Results are shown in the `comparison set 1' of Table \ref{table:comparison}. It shows that all the loading metrics and all the EQ scores are highly correlated with high significance level. 

\subsubsection{RQ2: Can the loading metrics effectively measure improvement of users' experience?}
It should be noted that the high correlation found in comparison set 1 of Table \ref{table:comparison} does not mean that the users' editing experience and the loading metrics are causally related. Editing experience may be related to other confounding factors such as the music content itself. To figure out whether the loading metrics truly affect users' experience of editing and which loading metrics can better reflect the subjective responses in the evaluation process, we explore the correlation between 1) the performance ratios of each transferred samples generated by the proposed and baseline methods, and 2) their differences of the responses in the subjective test questions. If a loading metric serves as an adequate descriptor of users' experience, then  \emph{the reduction of loading metric} and \emph{the improvement of users' response} should be strongly and significantly correlated. 
More specifically, we compute the Pearson's $r$ coefficient between each of the difference of the scores (denoted as $\Delta$EQ or $\Delta$LQ) between BMST and Baseline and performance ratio. The difference and ratio values are obtained from the two generated versions (BMST and Baseline) of each sample; such subtraction/ratio operations can exclude the confound of the original music contents. This idea of taking difference (or ratio) partly stems from the difference-in-differences (DID) estimation method \cite{bradley2020economics}.  



\begin{table}[t]
\centering
\begin{tabular}{|l||c c c|c|}
\hline
Comparison set 1 & EQ1 & EQ2 & EQ3 & LQ1 \\
 \hline
Time & -0.64$^{**}$ & -0.64$^{**}$ & -0.79$^{**}$ & -0.58$^{**}$ \\
Keyboard & -0.71$^{**}$ & -0.64$^{**}$ & -0.74$^{**}$ &  -0.68$^{**}$ \\
Mouse & -0.75$^{**}$ & -0.68$^{**}$ &	-0.83$^{**}$ & -0.69$^{**}$ \\
 \hline\hline
Comparison set 2 & $\Delta$EQ1 & $\Delta$EQ2 & $\Delta$EQ3 & $\Delta$LQ1 \\  
 \hline
 Time (ratio) & -0.58$^*$ & \textbf{-0.60$^*$} & -0.60$^*$ & -0.51 \\ 
 Keyboard (ratio) & \textbf{-0.73$^*$} & -0.55 & \textbf{-0.76$^{**}$} & -0.50 \\
 Mouse (ratio) & -0.56 & -0.51 & -0.57 & -0.54 \\
 \hline
 \hline
Comparison set 3 & EQ1 & EQ2 & EQ3 & LQ1 \\
 \hline
EQ1 & -- & 0.86$^{**}$ & 0.88$^{**}$ & 0.69$^{**}$\\
EQ2 & -- & -- & 0.85$^{**}$ & 0.61$^{**}$\\
EQ3 & -- & -- & -- & 0.69$^{**}$\\
 \hline\hline
Comparison set 4 & $\Delta$EQ1 & $\Delta$EQ2 & $\Delta$EQ3 & $\Delta$LQ1 \\  
 \hline
$\Delta$EQ1 & -- & 0.82$^{**}$ & 0.86$^{**}$ & 0.35 \\
$\Delta$EQ2 & -- & -- & 0.89$^{**}$ & 0.45 \\
$\Delta$EQ3 & -- & -- & -- & 0.58$^*$ \\
 \hline
\end{tabular}
\caption{Pearson's $r$ coefficients between the performance ratio and the improvement of users' subjective ratings on BMST comparing to the baseline. One star ($^*$): $p<0.05$. Double star ($^{**}$): $p<0.005$.}
\label{table:comparison}
\end{table}



\begin{figure}[t]
\centering
\includegraphics[width=0.75\linewidth, trim={0cm 0.2cm 0cm 0.2cm}, clip]{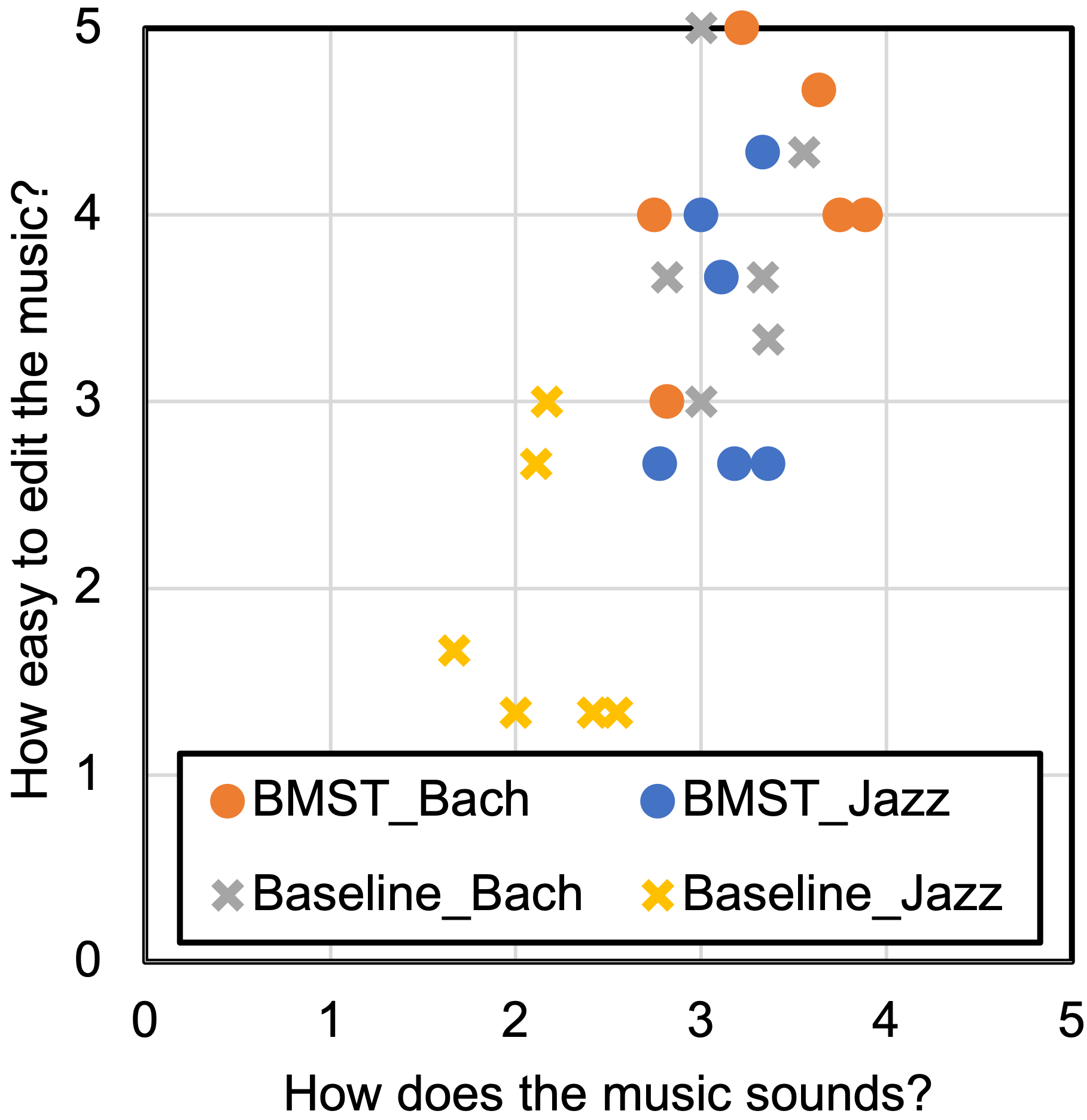}
\caption{Scattering plot of the listeners' and editors' responses on `how good the music sounds' (LQ1) and `how easy to edit the music' (EQ1) for the 24 samples.}
\label{fig: correlation}
\end{figure}

The comparison set 2 of Table \ref{table:comparison} lists the Pearson's $r$ coefficients between the performance ratios of the loading metrics, and the difference of ratings in the subjective questions (denoted as $\Delta$EQ1, $\Delta$EQ2, $\Delta$EQ3, and $\Delta$LQ1). Results show that the editing time and the amount of keyboard action are the two most important factors correlated to users' responses; $\Delta$EQ1 (easy to edit) and $\Delta$EQ3 (helpful for making music) are strongly correlated with the amount of keyboard press, whereas $\Delta$EQ2 (satisfactory result) are associated with the editing time. Interestingly, the amount of mouse click does not significantly reflect users' responses, probably because that key presses are usually applied to specific editing actions, 
while mouse clicks can be used for miscellaneous purposes (e.g., editing, selecting, or searching in a music clip), or just meaningless. 
With these discussions, we argue that correlation over the \emph{changes} of users' response and loading metrics rather than those values themselves can better explain the improvement/degradation of performance.




\subsubsection{(RQ3) What is the relationship between different aspects of users' editing experience?} 
The comparison set 3 of Table \ref{table:comparison} shows that EQ$i$ and EQ$j$, and $\Delta$EQ$i$ and $\Delta$EQ$j$, are both strongly correlated for $i,j\in\{1,2,3\}$, $i\neq j$. This means that ease of editing, satisfaction of edited result, and usefulness of the style transfer methods are strongly correlated to each other. Also, if a style transfer model can make users feel it easier and more satisfying in editing compared to a baseline model, then such improvement can be a measure of the improvement of the usefulness of the style transfer model.

\subsubsection{(RQ4) What is the relationship between users' listening and editing experience?}

The rightmost column of the comparison set 4 in Table \ref{table:comparison} shows that in comparison to other correlation, the changes of the editing test responses ($\Delta$EQ) are less correlated to the changes in the listening test results ($\Delta$LQ). To further investigate this phenomenon,
we plot the users' responses of 
EQ1 and LQ1 (the case with lowest correlation between $\Delta$EQ and $\Delta$LQ) in Figure \ref{fig: correlation}.  
Some detailed observations are worth mentioning: the two music samples that sounds the best (full LQ1 score) are not the easiest one to edit; there are around one-third of samples which sound better but harder to edit than them. 
On the other hand, the music samples that sound worst are almost surely the hardest ones to edit (mostly the baseline-to-Jazz cases).
The rock music pieces (e.g., \emph{Paranoid Android}) apply much more complicated rhythmic patterns and accompaniment arrangements than the kids folk songs, and such complexity results in several independent factors to challenge the learning of the model, the listening experience, and the editing process. This also explains why $\Delta$EQ1 and $\Delta$LQ1 correlate much weaker (Pearson's $r=0.35$, $p=0.26$) than EQ1 and LQ1 do: the EQ1-LQ1 plot confounds these independent factors together and exhibits a nice correlation as a result. This again explains the importance of performing correlation over difference. In summary, when excluding the influence from the original music content, we found that the editing test does provide profound insights, which cannot be explained by the listening test, and these aspects can be effectively described by the designed loading metrics such as the amount of key press and the editing time. 

\begin{figure}[t]
    \centering
    \includegraphics[width=0.87\linewidth]{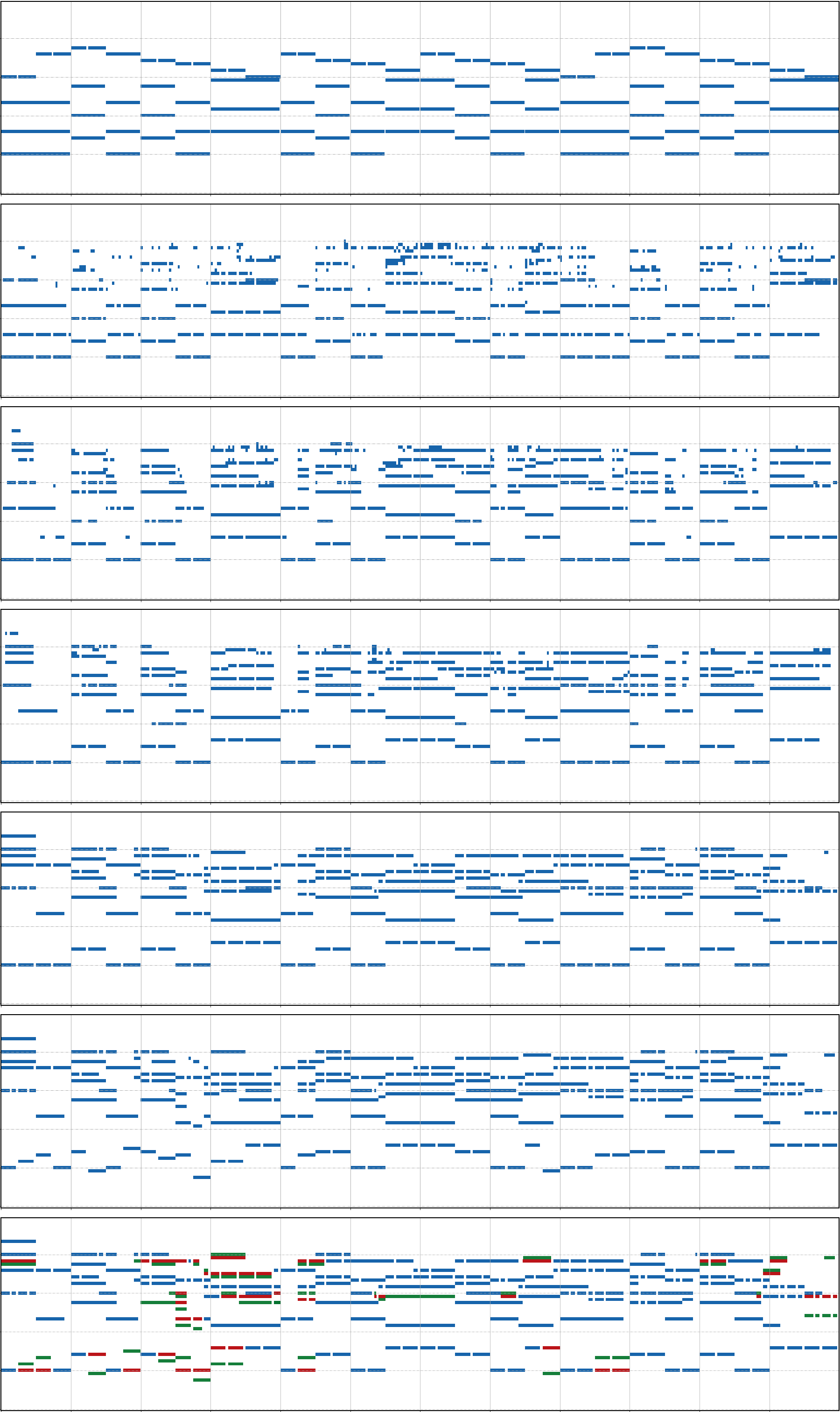}
    \caption{Piano rolls of the input, intermediate generated results, final result, and an example result after the editing test. From top to bottom: input piano roll, intermediate results at the 1st, 4th and 8th iterations of Gibbs sampling, the final output at the 15th iteration, the result after an editing test, and the difference between the model output and the editing result (green: inserted notes; red: deleted notes).}
    \label{fig:illustration}
\end{figure}

\subsection{Illustration of generated and edited results}

Figure \ref{fig:illustration} illustrates the input, intermediate outputs during Gibbs sampling, and final output of \emph{Twinkle Twinkle Little Stars} transferred to jazz style, accompanied with one of its edited output after the editing test. For the intermediate results, the outputs at the 1st, 4th, 8th, and 15th iterations (the final iteration) are depicted. We observe that in the initial step, a preliminary rhythmic pattern is created first, while the melody part remains unclear. In the 8th iteration, the bass line goes stable and the triad chords evolve to more complicated chords with richer rhythm. In the last two iterations, the superpositioned melody content stabilizes the whole result with a low annealing factor of Gibbs sampling. The illustrated edited result shows that the editing actions are mainly the enrichment of bass line and chord progression, which reveals the challenge of generating diverse music contents up to human's level.


\section{Concluding remarks}\label{sec:con}
The proposed bidirectional music style transfer model has demonstrated improved performance over the previous work in terms of the objective measure, listening test, and editing test. By correlation analysis over the difference of ratings and performance ratios, we indicate that the loading metrics can reflect users' edit experience, and the key press counts and editing time are more representative than the mouse clicks in assessing the quality of a music style transfer model. Also, listening tests and editing tests represents different aspects of users' experience; each of one cannot replace the other.  

The major issue of editing tests is that it requires more testing time and human resources than listening tests. Besides, one issue of the proposed style transfer method is also its inefficiency: in the inference stage, finishing 15 iterations of Gibbs sampling for a 16-bar music without batch processing takes around 1 hour using a V100 GPU, and the reason is that the iterative and non-chronological properties of Gibbs sampling make it behave like humans in editing music: ``human composer...scribbling motifs here and there, often revisiting choices previously made'' \cite{huang2019counterpoint}. These issues indicate the bottleneck to fulfill a human-in-the-loop music generation tool, 
and also the richness of information in humans' editing behaviors rather than listening ones. More in-depth user study, datasets and models focusing on the editing behaviors of music will be of central importance in future music generation research. 


\begin{acks}
This project was partly supported by MOST Taiwan and KKBOX Inc. under the project \emph{GenMusic Project: Industrial AI-Powered Music Composition Platform} (Grant No. 32T-1001212-2C). 
\end{acks}

\bibliographystyle{ACM-Reference-Format}
\balance
\bibliography{reference}


\begin{thebibliography}{41}


\ifx \showCODEN    \undefined \def \showCODEN     #1{\unskip}     \fi
\ifx \showDOI      \undefined \def \showDOI       #1{#1}\fi
\ifx \showISBNx    \undefined \def \showISBNx     #1{\unskip}     \fi
\ifx \showISBNxiii \undefined \def \showISBNxiii  #1{\unskip}     \fi
\ifx \showISSN     \undefined \def \showISSN      #1{\unskip}     \fi
\ifx \showLCCN     \undefined \def \showLCCN      #1{\unskip}     \fi
\ifx \shownote     \undefined \def \shownote      #1{#1}          \fi
\ifx \showarticletitle \undefined \def \showarticletitle #1{#1}   \fi
\ifx \showURL      \undefined \def \showURL       {\relax}        \fi
\providecommand\bibfield[2]{#2}
\providecommand\bibinfo[2]{#2}
\providecommand\natexlab[1]{#1}
\providecommand\showeprint[2][]{arXiv:#2}

\bibitem[\protect\citeauthoryear{Barnab{\`o}, Trappolini, Lastilla, Campagnano,
  Fan, Petroni, and Silvestri}{Barnab{\`o} et~al\mbox{.}}{2021}]%
        {barnabo2021cycledrums}
\bibfield{author}{\bibinfo{person}{Giorgio Barnab{\`o}},
  \bibinfo{person}{Giovanni Trappolini}, \bibinfo{person}{Lorenzo Lastilla},
  \bibinfo{person}{Cesare Campagnano}, \bibinfo{person}{Angela Fan},
  \bibinfo{person}{Fabio Petroni}, {and} \bibinfo{person}{Fabrizio Silvestri}.}
  \bibinfo{year}{2021}\natexlab{}.
\newblock \showarticletitle{CycleDRUMS: Automatic Drum Arrangement For Bass
  Lines Using CycleGAN}.
\newblock \bibinfo{journal}{\emph{arXiv preprint arXiv:2104.00353}}
  (\bibinfo{year}{2021}).
\newblock


\bibitem[\protect\citeauthoryear{Boulanger-Lewandowski, Bengio, and
  Vincent}{Boulanger-Lewandowski et~al\mbox{.}}{2012}]%
        {boulanger2012modeling}
\bibfield{author}{\bibinfo{person}{Nicolas Boulanger-Lewandowski},
  \bibinfo{person}{Yoshua Bengio}, {and} \bibinfo{person}{Pascal Vincent}.}
  \bibinfo{year}{2012}\natexlab{}.
\newblock \showarticletitle{Modeling temporal dependencies in high-dimensional
  sequences: Application to polyphonic music generation and transcription}. In
  \bibinfo{booktitle}{\emph{Proc. International Conference on Machine Learning
  (ICML)}}. \bibinfo{pages}{1881--1888}.
\newblock


\bibitem[\protect\citeauthoryear{Bradley and Green}{Bradley and Green}{2020}]%
        {bradley2020economics}
\bibfield{author}{\bibinfo{person}{Steve Bradley} {and} \bibinfo{person}{Colin
  Green}.} \bibinfo{year}{2020}\natexlab{}.
\newblock \showarticletitle{The Economics of Education: A Comprehensive
  Overview}.
\newblock  (\bibinfo{year}{2020}).
\newblock


\bibitem[\protect\citeauthoryear{Brunner, Wang, Wattenhofer, and Zhao}{Brunner
  et~al\mbox{.}}{2018}]%
        {brunner2018symbolic}
\bibfield{author}{\bibinfo{person}{Gino Brunner}, \bibinfo{person}{Yuyi Wang},
  \bibinfo{person}{Roger Wattenhofer}, {and} \bibinfo{person}{Sumu Zhao}.}
  \bibinfo{year}{2018}\natexlab{}.
\newblock \showarticletitle{Symbolic music genre transfer with cycle{GAN}}. In
  \bibinfo{booktitle}{\emph{Proceedings of the International Conference on
  Tools with Artificial Intelligence (ICTAI)}}. \bibinfo{pages}{786--793}.
\newblock


\bibitem[\protect\citeauthoryear{Chen and Su}{Chen and Su}{2019}]%
        {chen2019harmony}
\bibfield{author}{\bibinfo{person}{Tsung-Ping Chen} {and} \bibinfo{person}{Li
  Su}.} \bibinfo{year}{2019}\natexlab{}.
\newblock \showarticletitle{Harmony Transformer: Incorporating chord
  segmentation into harmony recognition}. In
  \bibinfo{booktitle}{\emph{Intenational Society of Music Information Retrieval
  Conference (ISMIR)}}. \bibinfo{pages}{259--267}.
\newblock


\bibitem[\protect\citeauthoryear{Choi, Hawthorne, Simon, Dinculescu, and
  Engel}{Choi et~al\mbox{.}}{2019}]%
        {choi2019encoding}
\bibfield{author}{\bibinfo{person}{Kristy Choi}, \bibinfo{person}{Curtis
  Hawthorne}, \bibinfo{person}{Ian Simon}, \bibinfo{person}{Monica Dinculescu},
  {and} \bibinfo{person}{Jesse Engel}.} \bibinfo{year}{2019}\natexlab{}.
\newblock \showarticletitle{Encoding musical style with transformer
  autoencoders}. In \bibinfo{booktitle}{\emph{International Conference on
  Machine Learning (ICML)}}. \bibinfo{pages}{1899--1908}.
\newblock


\bibitem[\protect\citeauthoryear{C{\' i}fka, {\c{S}}im{\c{s}}ekli, and
  Richard}{C{\' i}fka et~al\mbox{.}}{2019}]%
        {cifka2019supervised}
\bibfield{author}{\bibinfo{person}{Ond{\v{r}}ej C{\' i}fka},
  \bibinfo{person}{Umut {\c{S}}im{\c{s}}ekli}, {and} \bibinfo{person}{Ga{\" e}l
  Richard}.} \bibinfo{year}{2019}\natexlab{}.
\newblock \showarticletitle{Supervised Symbolic Music Style Translation Using
  Synthetic Data}. In \bibinfo{booktitle}{\emph{International Society of Music
  Information Retrieval Conference (ISMIR)}}. \bibinfo{pages}{588--595}.
\newblock


\bibitem[\protect\citeauthoryear{Colombo, Brea, and Gerstner}{Colombo
  et~al\mbox{.}}{2019}]%
        {colombo2018learning}
\bibfield{author}{\bibinfo{person}{Florian Colombo}, \bibinfo{person}{Johanni
  Brea}, {and} \bibinfo{person}{Wulfram Gerstner}.}
  \bibinfo{year}{2019}\natexlab{}.
\newblock \showarticletitle{Learning to Generate Music with BachProp}. In
  \bibinfo{booktitle}{\emph{Sound and Music Computing (SMC)}}.
\newblock


\bibitem[\protect\citeauthoryear{Cuthbert and Ariza}{Cuthbert and
  Ariza}{2010}]%
        {cuthbert2010music21}
\bibfield{author}{\bibinfo{person}{Michael~Scott Cuthbert} {and}
  \bibinfo{person}{Christopher Ariza}.} \bibinfo{year}{2010}\natexlab{}.
\newblock \showarticletitle{music21: A toolkit for computer-aided musicology
  and symbolic music data}. In \bibinfo{booktitle}{\emph{International Society
  for Music Information Retrieval Conference (ISMIR)}}.
\newblock


\bibitem[\protect\citeauthoryear{Dai and Xia}{Dai and Xia}{2018}]%
        {dai2018music}
\bibfield{author}{\bibinfo{person}{Shuqi Dai} {and} \bibinfo{person}{Gus Xia}.}
  \bibinfo{year}{2018}\natexlab{}.
\newblock \showarticletitle{Music Style Transfer Issues: A Position Paper}. In
  \bibinfo{booktitle}{\emph{the 6th International Workshop on Musical
  Metacreation (MUME)}}.
\newblock


\bibitem[\protect\citeauthoryear{Devlin, Chang, Lee, and Toutanova}{Devlin
  et~al\mbox{.}}{2018}]%
        {devlin2018bert}
\bibfield{author}{\bibinfo{person}{Jacob Devlin}, \bibinfo{person}{Ming-Wei
  Chang}, \bibinfo{person}{Kenton Lee}, {and} \bibinfo{person}{Kristina
  Toutanova}.} \bibinfo{year}{2018}\natexlab{}.
\newblock \showarticletitle{Bert: Pre-training of deep bidirectional
  transformers for language understanding}. In
  \bibinfo{booktitle}{\emph{Proceedings of NAACL-HLT}}.
\newblock


\bibitem[\protect\citeauthoryear{Dhariwal, Jun, Payne, Kim, Radford, and
  Sutskever}{Dhariwal et~al\mbox{.}}{2020}]%
        {dhariwal2020jukebox}
\bibfield{author}{\bibinfo{person}{Prafulla Dhariwal}, \bibinfo{person}{Heewoo
  Jun}, \bibinfo{person}{Christine Payne}, \bibinfo{person}{Jong~Wook Kim},
  \bibinfo{person}{Alec Radford}, {and} \bibinfo{person}{Ilya Sutskever}.}
  \bibinfo{year}{2020}\natexlab{}.
\newblock \showarticletitle{Jukebox: A generative model for music}.
\newblock \bibinfo{journal}{\emph{arXiv preprint arXiv:2005.00341}}
  (\bibinfo{year}{2020}).
\newblock


\bibitem[\protect\citeauthoryear{Dong, Hsiao, Yang, and Yang}{Dong
  et~al\mbox{.}}{2017}]%
        {dong2017musegan}
\bibfield{author}{\bibinfo{person}{Hao-Wen Dong}, \bibinfo{person}{Wen-Yi
  Hsiao}, \bibinfo{person}{Li-Chia Yang}, {and} \bibinfo{person}{Yi-Hsuan
  Yang}.} \bibinfo{year}{2017}\natexlab{}.
\newblock \showarticletitle{MuseGAN: Symbolic-domain music generation and
  accompaniment with multi-track sequential generative adversarial networks}.
  In \bibinfo{booktitle}{\emph{Thirty-Second AAAI Conference on Artificial
  Intelligence (AAAI)}}.
\newblock


\bibitem[\protect\citeauthoryear{Engel, Resnick, Roberts, Dieleman, Norouzi,
  Eck, and Simonyan}{Engel et~al\mbox{.}}{2017}]%
        {engel2017neural}
\bibfield{author}{\bibinfo{person}{Jesse Engel}, \bibinfo{person}{Cinjon
  Resnick}, \bibinfo{person}{Adam Roberts}, \bibinfo{person}{Sander Dieleman},
  \bibinfo{person}{Mohammad Norouzi}, \bibinfo{person}{Douglas Eck}, {and}
  \bibinfo{person}{Karen Simonyan}.} \bibinfo{year}{2017}\natexlab{}.
\newblock \showarticletitle{Neural audio synthesis of musical notes with
  wavenet autoencoders}. In \bibinfo{booktitle}{\emph{Proceedings of the
  International Conference on Machine Learning (ICML)}}.
  \bibinfo{pages}{1068--1077}.
\newblock


\bibitem[\protect\citeauthoryear{Ens and Pasquier}{Ens and Pasquier}{2020}]%
        {ens2020quantifying}
\bibfield{author}{\bibinfo{person}{Jeff Ens} {and} \bibinfo{person}{Philippe
  Pasquier}.} \bibinfo{year}{2020}\natexlab{}.
\newblock \showarticletitle{Quantifying Musical Style: Ranking Symbolic Music
  based on Similarity to a Style}. In \bibinfo{booktitle}{\emph{Intenational
  Society of Music Information Retrieval Conference (ISMIR)}}.
  \bibinfo{pages}{870--877}.
\newblock


\bibitem[\protect\citeauthoryear{Hadjeres, Pachet, and Nielsen}{Hadjeres
  et~al\mbox{.}}{2017}]%
        {hadjeres2017deepbach}
\bibfield{author}{\bibinfo{person}{Ga{\"e}tan Hadjeres},
  \bibinfo{person}{Fran{\c{c}}ois Pachet}, {and} \bibinfo{person}{Frank
  Nielsen}.} \bibinfo{year}{2017}\natexlab{}.
\newblock \showarticletitle{Deepbach: a steerable model for bach chorales
  generation}. In \bibinfo{booktitle}{\emph{Proceedings of the International
  Conference on Machine Learning (ICML)}}. JMLR. org,
  \bibinfo{pages}{1362--1371}.
\newblock


\bibitem[\protect\citeauthoryear{Hawthorne, Huang, Ippolito, and Eck}{Hawthorne
  et~al\mbox{.}}{2018}]%
        {hawthorne2018transformer}
\bibfield{author}{\bibinfo{person}{Curtis Hawthorne}, \bibinfo{person}{Anna
  Huang}, \bibinfo{person}{Daphne Ippolito}, {and} \bibinfo{person}{Douglas
  Eck}.} \bibinfo{year}{2018}\natexlab{}.
\newblock \showarticletitle{Transformer-{NADE} for Piano Performances}. In
  \bibinfo{booktitle}{\emph{NIPS 2nd Workshop on Machine Learning for
  Creativity and Design}}.
\newblock


\bibitem[\protect\citeauthoryear{Holzapfel, Benetos, et~al\mbox{.}}{Holzapfel
  et~al\mbox{.}}{2019}]%
        {holzapfel2019automatic}
\bibfield{author}{\bibinfo{person}{Andre Holzapfel}, \bibinfo{person}{Emmanouil
  Benetos}, {et~al\mbox{.}}} \bibinfo{year}{2019}\natexlab{}.
\newblock \showarticletitle{Automatic music transcription and ethnomusicology:
  a user study}. In \bibinfo{booktitle}{\emph{Proceedings of the International
  Society for Music Information Retrieval Conference (ISMIR)}}.
  \bibinfo{pages}{678--684}.
\newblock


\bibitem[\protect\citeauthoryear{Huang, Cooijmans, Roberts, Courville, and
  Eck}{Huang et~al\mbox{.}}{2017}]%
        {huang2019counterpoint}
\bibfield{author}{\bibinfo{person}{Cheng-Zhi~Anna Huang}, \bibinfo{person}{Tim
  Cooijmans}, \bibinfo{person}{Adam Roberts}, \bibinfo{person}{Aaron
  Courville}, {and} \bibinfo{person}{Douglas Eck}.}
  \bibinfo{year}{2017}\natexlab{}.
\newblock \showarticletitle{Counterpoint by convolution}. In
  \bibinfo{booktitle}{\emph{International Society of Music Information
  Retrieval Conference (ISMIR)}}.
\newblock


\bibitem[\protect\citeauthoryear{Huang, Koops, Newton-Rex, Dinculescu, and
  Cai}{Huang et~al\mbox{.}}{2020}]%
        {huang2020ai}
\bibfield{author}{\bibinfo{person}{Cheng-Zhi~Anna Huang},
  \bibinfo{person}{Hendrik~Vincent Koops}, \bibinfo{person}{Ed Newton-Rex},
  \bibinfo{person}{Monica Dinculescu}, {and} \bibinfo{person}{Carrie~J. Cai}.}
  \bibinfo{year}{2020}\natexlab{}.
\newblock \showarticletitle{Human-AI Co-Creation in Songwriting}. In
  \bibinfo{booktitle}{\emph{Intenational Society of Music Information Retrieval
  Conference (ISMIR)}}. \bibinfo{pages}{708--716}.
\newblock


\bibitem[\protect\citeauthoryear{Huang, Vaswani, Uszkoreit, Simon, Hawthorne,
  Shazeer, Dai, Hoffman, Dinculescu, and Eck}{Huang et~al\mbox{.}}{2019}]%
        {huang2018music}
\bibfield{author}{\bibinfo{person}{Cheng-Zhi~Anna Huang},
  \bibinfo{person}{Ashish Vaswani}, \bibinfo{person}{Jakob Uszkoreit},
  \bibinfo{person}{Ian Simon}, \bibinfo{person}{Curtis Hawthorne},
  \bibinfo{person}{Noam Shazeer}, \bibinfo{person}{Andrew~M. Dai},
  \bibinfo{person}{Matthew~D. Hoffman}, \bibinfo{person}{Monica Dinculescu},
  {and} \bibinfo{person}{Douglas Eck}.} \bibinfo{year}{2019}\natexlab{}.
\newblock \showarticletitle{Music transformer: Generating music with long-term
  structure}. In \bibinfo{booktitle}{\emph{International Conference on Learning
  Representations (ICLR)}}.
\newblock


\bibitem[\protect\citeauthoryear{Huang and Yang}{Huang and Yang}{2020}]%
        {huang2020pop}
\bibfield{author}{\bibinfo{person}{Yu-Siang Huang} {and}
  \bibinfo{person}{Yi-Hsuan Yang}.} \bibinfo{year}{2020}\natexlab{}.
\newblock \showarticletitle{Pop Music Transformer: Generating Music with Rhythm
  and Harmony}. In \bibinfo{booktitle}{\emph{ACM International Conference on
  Multimedia (ACM MM)}}. \bibinfo{pages}{1180–--1188}.
\newblock


\bibitem[\protect\citeauthoryear{Liang, Gotham, Johnson, and Shotton}{Liang
  et~al\mbox{.}}{2017}]%
        {liang2017automatic}
\bibfield{author}{\bibinfo{person}{Feynman~T. Liang}, \bibinfo{person}{Mark
  Gotham}, \bibinfo{person}{Matthew Johnson}, {and} \bibinfo{person}{Jamie
  Shotton}.} \bibinfo{year}{2017}\natexlab{}.
\newblock \showarticletitle{Automatic Stylistic Composition of Bach Chorales
  with Deep LSTM.}. In \bibinfo{booktitle}{\emph{International Society of Music
  Information Retrieval Conference (ISMIR)}}. \bibinfo{pages}{449--456}.
\newblock


\bibitem[\protect\citeauthoryear{Lin, Goyal, Girshick, He, and Doll{\'a}r}{Lin
  et~al\mbox{.}}{2017}]%
        {lin2017focal}
\bibfield{author}{\bibinfo{person}{Tsung-Yi Lin}, \bibinfo{person}{Priya
  Goyal}, \bibinfo{person}{Ross Girshick}, \bibinfo{person}{Kaiming He}, {and}
  \bibinfo{person}{Piotr Doll{\'a}r}.} \bibinfo{year}{2017}\natexlab{}.
\newblock \showarticletitle{Focal loss for dense object detection}. In
  \bibinfo{booktitle}{\emph{Proceedings of the IEEE International Conference on
  Computer Vision (ICCV)}}. \bibinfo{pages}{2980--2988}.
\newblock


\bibitem[\protect\citeauthoryear{Lu, Xue, Chang, Lee, and Su}{Lu
  et~al\mbox{.}}{2019}]%
        {lu2018play}
\bibfield{author}{\bibinfo{person}{Chien-Yu Lu}, \bibinfo{person}{Min-Xin Xue},
  \bibinfo{person}{Chia-Che Chang}, \bibinfo{person}{Che-Rung Lee}, {and}
  \bibinfo{person}{Li Su}.} \bibinfo{year}{2019}\natexlab{}.
\newblock \showarticletitle{Play as you like: Timbre-enhanced multi-modal music
  style transfer}. In \bibinfo{booktitle}{\emph{Proceedings of the AAAI
  Conference on Artificial Intelligence (AAAI)}}, Vol.~\bibinfo{volume}{33}.
  \bibinfo{pages}{1061--1068}.
\newblock


\bibitem[\protect\citeauthoryear{Lu and Su}{Lu and Su}{2018}]%
        {lu2018transferring}
\bibfield{author}{\bibinfo{person}{Wei~Tsung Lu} {and} \bibinfo{person}{Li
  Su}.} \bibinfo{year}{2018}\natexlab{}.
\newblock \showarticletitle{Transferring the Style of Homophonic Music Using
  Recurrent Neural Networks and Autoregressive Model.}. In
  \bibinfo{booktitle}{\emph{Proceedings of the International Society for Music
  Information Retrieval Conference (ISMIR)}}. \bibinfo{pages}{740--746}.
\newblock


\bibitem[\protect\citeauthoryear{Maezawa}{Maezawa}{2018}]%
        {maezawa2018deep}
\bibfield{author}{\bibinfo{person}{Akira Maezawa}.}
  \bibinfo{year}{2018}\natexlab{}.
\newblock \showarticletitle{Deep piano performance rendering with conditional
  {VAE}}. In \bibinfo{booktitle}{\emph{ISMIR Late Breaking and Demo Papers}}.
\newblock


\bibitem[\protect\citeauthoryear{Malik and Ek}{Malik and Ek}{2017}]%
        {malik2017neural}
\bibfield{author}{\bibinfo{person}{Iman Malik} {and}
  \bibinfo{person}{Carl~Henrik Ek}.} \bibinfo{year}{2017}\natexlab{}.
\newblock \showarticletitle{Neural translation of musical style}.
\newblock \bibinfo{journal}{\emph{arXiv preprint arXiv:1708.03535}}
  (\bibinfo{year}{2017}).
\newblock


\bibitem[\protect\citeauthoryear{Mauch, Cannam, Bittner, Fazekas, Salamon, Dai,
  Bello, and Dixon}{Mauch et~al\mbox{.}}{2015}]%
        {mauch2015computer}
\bibfield{author}{\bibinfo{person}{Matthias Mauch}, \bibinfo{person}{Chris
  Cannam}, \bibinfo{person}{Rachel Bittner}, \bibinfo{person}{George Fazekas},
  \bibinfo{person}{Justin Salamon}, \bibinfo{person}{Jaijie Dai},
  \bibinfo{person}{Juan Bello}, {and} \bibinfo{person}{Simon Dixon}.}
  \bibinfo{year}{2015}\natexlab{}.
\newblock \showarticletitle{Computer-aided melody note transcription using the
  tony software: Accuracy and efficiency}. In \bibinfo{booktitle}{\emph{Proc.
  Sound and Music Computing (SMC)}}.
\newblock


\bibitem[\protect\citeauthoryear{Oord, Dieleman, Zen, Simonyan, Vinyals,
  Graves, Kalchbrenner, Senior, and Kavukcuoglu}{Oord et~al\mbox{.}}{2016}]%
        {oord2016wavenet}
\bibfield{author}{\bibinfo{person}{Aaron van~den Oord}, \bibinfo{person}{Sander
  Dieleman}, \bibinfo{person}{Heiga Zen}, \bibinfo{person}{Karen Simonyan},
  \bibinfo{person}{Oriol Vinyals}, \bibinfo{person}{Alex Graves},
  \bibinfo{person}{Nal Kalchbrenner}, \bibinfo{person}{Andrew Senior}, {and}
  \bibinfo{person}{Koray Kavukcuoglu}.} \bibinfo{year}{2016}\natexlab{}.
\newblock \showarticletitle{Wavenet: A generative model for raw audio}.
\newblock \bibinfo{journal}{\emph{arXiv preprint arXiv:1609.03499}}
  (\bibinfo{year}{2016}).
\newblock


\bibitem[\protect\citeauthoryear{Parisotto, Song, Rae, Pascanu, Gulcehre,
  Jayakumar, Jaderberg, Kaufman, Clark, Noury, et~al\mbox{.}}{Parisotto
  et~al\mbox{.}}{2020}]%
        {parisotto2019stabilizing}
\bibfield{author}{\bibinfo{person}{Emilio Parisotto},
  \bibinfo{person}{H~Francis Song}, \bibinfo{person}{Jack~W Rae},
  \bibinfo{person}{Razvan Pascanu}, \bibinfo{person}{Caglar Gulcehre},
  \bibinfo{person}{Siddhant~M Jayakumar}, \bibinfo{person}{Max Jaderberg},
  \bibinfo{person}{Raphael~Lopez Kaufman}, \bibinfo{person}{Aidan Clark},
  \bibinfo{person}{Seb Noury}, {et~al\mbox{.}}}
  \bibinfo{year}{2020}\natexlab{}.
\newblock \showarticletitle{Stabilizing Transformers for Reinforcement
  Learning}. In \bibinfo{booktitle}{\emph{International Conference on Machine
  Learning (ICML)}}. \bibinfo{pages}{7487--7498}.
\newblock


\bibitem[\protect\citeauthoryear{Roberts, Engel, Raffel, Hawthorne, and
  Eck}{Roberts et~al\mbox{.}}{2018}]%
        {roberts2018hierarchical}
\bibfield{author}{\bibinfo{person}{Adam Roberts}, \bibinfo{person}{Jesse
  Engel}, \bibinfo{person}{Colin Raffel}, \bibinfo{person}{Curtis Hawthorne},
  {and} \bibinfo{person}{Douglas Eck}.} \bibinfo{year}{2018}\natexlab{}.
\newblock \showarticletitle{A hierarchical latent vector model for learning
  long-term structure in music}. In \bibinfo{booktitle}{\emph{International
  Conference on Machine Learning (ICML)}}. PMLR, \bibinfo{pages}{4364--4373}.
\newblock


\bibitem[\protect\citeauthoryear{Sturm, Santos, Ben-Tal, and Korshunova}{Sturm
  et~al\mbox{.}}{2016}]%
        {sturm2016music}
\bibfield{author}{\bibinfo{person}{Bob Sturm}, \bibinfo{person}{Jo{\~a}o~Felipe
  Santos}, \bibinfo{person}{Oded Ben-Tal}, {and} \bibinfo{person}{Iryna
  Korshunova}.} \bibinfo{year}{2016}\natexlab{}.
\newblock \showarticletitle{Music Transcription Modelling and Composition Using
  Deep Learning}. In \bibinfo{booktitle}{\emph{1st Conference on Computer
  Simulation of Musical Creativity (CSMC)}}.
\newblock


\bibitem[\protect\citeauthoryear{Van~Oord, Kalchbrenner, and
  Kavukcuoglu}{Van~Oord et~al\mbox{.}}{2016}]%
        {oord2016pixel}
\bibfield{author}{\bibinfo{person}{Aaron Van~Oord}, \bibinfo{person}{Nal
  Kalchbrenner}, {and} \bibinfo{person}{Koray Kavukcuoglu}.}
  \bibinfo{year}{2016}\natexlab{}.
\newblock \showarticletitle{Pixel recurrent neural networks}. In
  \bibinfo{booktitle}{\emph{International Conference on Machine Learning
  (ICML)}}. PMLR, \bibinfo{pages}{1747--1756}.
\newblock


\bibitem[\protect\citeauthoryear{Vaswani, Shazeer, Parmar, Uszkoreit, Jones,
  Gomez, Kaiser, and Polosukhin}{Vaswani et~al\mbox{.}}{2017}]%
        {vaswani2017attention}
\bibfield{author}{\bibinfo{person}{Ashish Vaswani}, \bibinfo{person}{Noam
  Shazeer}, \bibinfo{person}{Niki Parmar}, \bibinfo{person}{Jakob Uszkoreit},
  \bibinfo{person}{Llion Jones}, \bibinfo{person}{Aidan~N Gomez},
  \bibinfo{person}{{\L}ukasz Kaiser}, {and} \bibinfo{person}{Illia
  Polosukhin}.} \bibinfo{year}{2017}\natexlab{}.
\newblock \showarticletitle{Attention is all you need}. In
  \bibinfo{booktitle}{\emph{Advances in Neural Information Processing Systems
  (NIPS)}}. \bibinfo{pages}{5998--6008}.
\newblock


\bibitem[\protect\citeauthoryear{Verma and Smith}{Verma and Smith}{2018}]%
        {verma2018neural}
\bibfield{author}{\bibinfo{person}{Prateek Verma} {and}
  \bibinfo{person}{Julius~O. Smith}.} \bibinfo{year}{2018}\natexlab{}.
\newblock \showarticletitle{Neural style transfer for audio spectograms}.
\newblock \bibinfo{journal}{\emph{arXiv preprint arXiv:1801.01589}}
  (\bibinfo{year}{2018}).
\newblock


\bibitem[\protect\citeauthoryear{Vogl, Leimeister, Nuan{\'a}in, Hlatky,
  Jord{\`a}~Puig, and Knees}{Vogl et~al\mbox{.}}{2016}]%
        {vogl2016intelligent}
\bibfield{author}{\bibinfo{person}{Richard Vogl}, \bibinfo{person}{Matthias
  Leimeister}, \bibinfo{person}{C{\'a}rthach~{\'O} Nuan{\'a}in},
  \bibinfo{person}{Michael Hlatky}, \bibinfo{person}{Sergi Jord{\`a}~Puig},
  {and} \bibinfo{person}{Peter Knees}.} \bibinfo{year}{2016}\natexlab{}.
\newblock \showarticletitle{An intelligent interface for drum pattern variation
  and comparative evaluation of algorithms}.
\newblock \bibinfo{journal}{\emph{Journal of the Audio Engineering Society}}
  \bibinfo{volume}{64}, \bibinfo{number}{7/8} (\bibinfo{year}{2016}),
  \bibinfo{pages}{503--513}.
\newblock


\bibitem[\protect\citeauthoryear{Wang, Li, and Smola}{Wang
  et~al\mbox{.}}{2019}]%
        {wang2019language}
\bibfield{author}{\bibinfo{person}{Chenguang Wang}, \bibinfo{person}{Mu Li},
  {and} \bibinfo{person}{Alexander~J Smola}.} \bibinfo{year}{2019}\natexlab{}.
\newblock \showarticletitle{Language models with transformers}.
\newblock \bibinfo{journal}{\emph{arXiv preprint arXiv:1904.09408}}
  (\bibinfo{year}{2019}).
\newblock


\bibitem[\protect\citeauthoryear{Wei, Wu, and Su}{Wei et~al\mbox{.}}{2019}]%
        {wei2019generating}
\bibfield{author}{\bibinfo{person}{I-Chieh Wei}, \bibinfo{person}{Chih-Wei Wu},
  {and} \bibinfo{person}{Li Su}.} \bibinfo{year}{2019}\natexlab{}.
\newblock \showarticletitle{Generating Structured Drum Pattern Using
  Variational Autoencoder and Self-similarity Matrix.}. In
  \bibinfo{booktitle}{\emph{Intenational Society of Music Information Retrieval
  Conference (ISMIR)}}. \bibinfo{pages}{847--854}.
\newblock


\bibitem[\protect\citeauthoryear{Yang and Lerch}{Yang and Lerch}{2020}]%
        {yang2020evaluation}
\bibfield{author}{\bibinfo{person}{Li-Chia Yang} {and}
  \bibinfo{person}{Alexander Lerch}.} \bibinfo{year}{2020}\natexlab{}.
\newblock \showarticletitle{On the evaluation of generative models in music}.
\newblock \bibinfo{journal}{\emph{Neural Computing and Applications}}
  \bibinfo{volume}{32}, \bibinfo{number}{9} (\bibinfo{year}{2020}),
  \bibinfo{pages}{4773--4784}.
\newblock


\bibitem[\protect\citeauthoryear{Yeh, Hsiao, Fukayama, Kitahara, Genchel, Liu,
  Dong, Chen, Leong, and Yang}{Yeh et~al\mbox{.}}{2020}]%
        {yeh2020automatic}
\bibfield{author}{\bibinfo{person}{Yin-Cheng Yeh}, \bibinfo{person}{Wen-Yi
  Hsiao}, \bibinfo{person}{Satoru Fukayama}, \bibinfo{person}{Tetsuro
  Kitahara}, \bibinfo{person}{Benjamin Genchel}, \bibinfo{person}{Hao-Min Liu},
  \bibinfo{person}{Hao-Wen Dong}, \bibinfo{person}{Yian Chen},
  \bibinfo{person}{Terence Leong}, {and} \bibinfo{person}{Yi-Hsuan Yang}.}
  \bibinfo{year}{2020}\natexlab{}.
\newblock \showarticletitle{Automatic Melody Harmonization with Triad Chords: A
  Comparative Study}.
\newblock \bibinfo{journal}{\emph{Journal of New Music Research (JNMR)}}
  \bibinfo{volume}{50}, \bibinfo{number}{1} (\bibinfo{year}{2020}),
  \bibinfo{pages}{37--51}.
\newblock


\end{thebibliography}







\end{document}